\documentclass[journal,twoside,web]{ieeecolor}
\usepackage{generic}
\PassOptionsToPackage{svgnames}{xcolor} 
\usepackage{textcomp}
\usepackage{graphicx} 
\graphicspath{{figure/}}
\usepackage{float}

\let\labelindent\relax
\usepackage{amsmath,amssymb,amsthm,bm}

\usepackage{array}
\usepackage[caption=false,font=footnotesize]{subfig}
\usepackage{cite}

\usepackage{algorithm,algorithmic}

\usepackage{bookmark}

\usepackage{bbm}
\usepackage{amsfonts}
\usepackage{array}

\usepackage{dsfont}
\usepackage{mathtools}
\usepackage{siunitx}

\usepackage{enumitem}

\DeclareSIUnit[]{\pu}{p.u.}
\DeclareSIUnit[]{\VA}{VA}

\DeclareSymbolFont{bbold}{U}{bbold}{m}{n}
\DeclareSymbolFontAlphabet{\mathbbold}{bbold}

\newcommand{\range}[1]{\ensuremath{\mathrm{range}(#1)}}

\DeclarePairedDelimiterX\Set[2]{\lbrace}{\rbrace}%
{ #1 \,\delimsize| \,\mathopen{} #2 }

\newcommand{\real}[0]{\mathbb R}

\usepackage{tikz}
\newcommand*\circled[1]{\tikz[baseline=(char.base)]{\node[shape=circle,draw,inner sep=0.05pt] (char) {#1};}}

\newtheorem{theorem}{Theorem}

\newtheorem{remark}{Remark}
\newtheorem{lemma}{Lemma}
\newtheorem{assumption}{Assumption}
\newtheorem{definition}{Definition}

\newtheorem{design}{Controller Design}
\newtheorem{example}{Example}

\allowdisplaybreaks

\def\BibTeX{{\rm B\kern-.05em{\sc i\kern-.025em b}\kern-.08em
    T\kern-.1667em\lower.7ex\hbox{E}\kern-.125emX}}
\begin{document}
\bibliographystyle{IEEEtran}

\title{Structured Neural-PI Control for Networked Systems: Stability and Steady-State Optimality Guarantees}
\author{Wenqi Cui, Yan Jiang,  Baosen Zhang, and Yuanyuan Shi
\thanks{Wenqi Cui, Yan Jiang, and Baosen Zhang are with the Department
of Electrical and Computer Engineering, University of Washington, Seattle, WA, 98195 e-mails:\{wenqicui, jiangyan, zhangbao\}@uw.edu}%
\thanks{Yuanyuan Shi is with the Department of Electrical and Computer Engineering, University of California San Diego e-mail: yyshi@eng.ucsd.edu} 
\thanks{The authors are supported in part by the National Science Foundation grant ECCS-1930605, ECCS-1942326, ECCS-2200692, ECCS-2153937 and the Washington Clean Energy Institute.}}


\maketitle

\begin{abstract}
 We study the control of networked systems with the goal of optimizing both transient and steady-state performances while providing stability guarantees. Linear proportional-integral (PI) controllers are almost always used in practice, but the linear parameterization of the controller fundamentally limits its performance. Learning-based approaches are becoming popular in designing nonlinear controllers, but the lack of stability guarantees makes the learned controllers difficult to apply in practical applications.
  This paper bridges the gap between neural network-based controller design and the need for stability guarantees. Using equilibrium-independent passivity, a property present in a wide range of physical systems, we propose structured neural-PI controllers that have provable guarantees on the convergence of output to a desired agreement value. If communication between neighbours is available, we further extend the controller to distributedly achieve optimal resource allocation at the steady state. We explicitly characterize the stability conditions and engineer neural networks that satisfy them by design. Experiments on traffic and power networks demonstrate that the proposed approach can improve transient and steady-state performances compared to existing state-of-the-art, while unstructured neural networks lead to unstable behaviors. 
\end{abstract}




\section{Introduction}\label{sec: intro}
We study the control of networked physical systems, where a large number of individual subsystems are connected to each other and operate in tandem. Such networked systems are present in numerous applications, and understanding their behaviors has been an active line of research~\cite{arcak2016networks,coogan2014dissipativity,simpson2018equilibrium}.
Currently, much of the effort has focused on characterizing properties of the subsystems and their interconnections to certify system stability. At the same time, the performance of these systems, that is, their ability to achieve certain objectives, is also of critical importance. 
However, it is challenging to optimize system performance using existing analytical approaches.

This paper aims to establish a framework for controller design that not only guarantees stability, but also  optimizes the performance in both the transient and steady-state period.
In particular, we consider networked systems where the output of subsystems need to reach an agreement at the steady state~\cite{burger2014duality,dorfler2014synchronization}. For example, vehicles in a platoon need to reach the same velocity~\cite{coogan2014dissipativity}  and generators in a power grid need to reach the same rotational speed~\cite{cui2022tps}.
This steady state should be reached quickly without causing too large state deviations and control effort during the transient period, making optimization of the transient performance important. Once the system reaches the steady state, we want to select the control action that achieves output agreement at the lowest cost.

For the system without real-time computation capability, linear Proportional-Integral (PI)  controllers are almost always used in practice to achieve the above goals~\cite{Zhao2015acc, andreasson2014distributed,  weitenberg2018exponential, SCHIFFER2017auto, golnaraghi2017automatic}. 
A proportional term provides instant feedback to improve the transient performance and an integral term drives the system outputs to the desired value at the steady state~\cite{Zhao2015acc, andreasson2014distributed}. 
If communication is available, previous works~\cite{Zhao2015acc, andreasson2014distributed,  weitenberg2018exponential, SCHIFFER2017auto} also tailor the integral term to realize distributed optimal resource allocation under quadratic costs. A linear parameterization, however, fundamentally restricts the degrees of freedom of a controller and can lead to suboptimal performances.


In the last decade, learning-based approaches become popular for nonlinear controller optimization over different objective functions~\cite{zhang2021multi, qu2020scalable, wang2021multi}. The nonlinear control law is normally represented through neural networks, which can then be optimized through learning algorithms.  Many  works show that learning-based approaches improve the performance by  at least 10\%-50\%  compared with conventional linear approaches~\cite{he2020reinforcement, yuan2022reinforcement, wang2021multi, cui2022tps}.
However, providing provable guarantees on stability and steady-state optimality for neural network-based controllers has been challenging. 

On the one hand, it is nontrivial to certify \textit{stability} of a learning-based control for all the possible initial states. Most works require a linear system model, and sometimes also a linear controller, such as LQR problems and its variants~\cite{lewis2009reinforcement, fazel2018global, lisafe}. 
The stability for nonlinear system model is more difficult. Many works use soft penalty on the violation of stability conditions in the cost function, but it cannot guarantee stability for all the possible initial states~\cite{chang2019neural}. 
Our previous works in~\cite{ cui2022tps,shi2021stability, jiang2022stable} show that stability of learning-based controllers can be enforced through special design of neural networks, but they rely on tailor-made Lyapunov functions and are limited to specific applications.

On the other hand, current learning-based approaches generally only optimize transient performance but neglect the \emph{steady-state optimality}. Even when training with long time horizons is computationally feasible, it is difficult to quantify how long is enough to reach the steady state, thus making steady-state performance optimization difficult. In contrast, linear PI controllers can achieve steady-state optimality for a large range of systems but may result in sub-optimal transient period cost, especially when the system dynamics and cost functions are nonlinear~\cite{he2020reinforcement,yuan2022reinforcement}. 
This work addresses the following open question:
\textit{Can we design learning-based controllers for a generalized class of networked systems, that optimize both the transient and steady-state performance, while guaranteeing system stability?}

Clearly, it is not possible to design a controller for everything and the answer depends on picking the right abstraction of the system. Passivity is a classical notion in control theory to characterize the inherent property of dynamical
systems by how their inputs and outputs correlate~\cite{arcak2016networks,hines2011equilibrium}. Many systems have been shown to be equilibrium-independent
passive (EIP)~\cite{burger2014duality,cui2022equilibrium,meissen2017passivity}, which characterizes passivity referenced to an arbitrary equilibrium input/output pair. This abstraction allows us to design generalized controllers for networked systems without considering their detailed dynamics. 

\textbf{Contributions.} This paper focuses on controller design for networked systems where the node dynamics are EIP. We propose a structured neural-PI controller that has provable stability guarantees and achieves steady-state optimal resource allocation. The key structure we use are monotonically increasing functions, and they are parameterized by what we call \emph{monotone neural networks}. We explicitly characterize the structural conditions of monotone neural networks and prove their universal approximation capability for monotonic functions. This way, transient performances can be optimized by the training of monotone neural networks, while stability and steady-state optimality are inherently guaranteed by design. We summarize contributions as follows.
\begin{enumerate}[noitemsep,nolistsep,leftmargin=*, label=\arabic*)]
    \item We construct a framework for neural network-based controller design that optimizes both the transient and steady-state performance of networked systems. We adopt a modular approach of stability analysis based on equilibrium-independent passivity, making the framework scalable to large systems and also robust to network topology and parameter variations.
    \item For networked systems without communication, we propose a neural-PI control law (Controller Design \ref{design: agreement_set}) that can be implemented fully decentralizedly with only local information. We prove that the controller design guarantees the convergence of output to a desired agreement value.
    \item For networked systems with communication, we propose a neural-PI control law (Controller Design \ref{design: optimal_flow}) where neighbouring nodes can exchange information of their marginal costs. We prove that this design guarantees both transient stability and steady-state optimality for a range of objective functions that include, but is not restricted to, quadratic cost functions. 
    \item Experiments on the control of vehicle platoons and power networks demonstrate that the proposed approach can reduce the transient cost by at least 30\% compared to optimized
    linear controllers, ensure stability and obtain optimal steady-state cost when communication is available. Unstructured neural networks, on the other hand, often lead to unstable behaviors. 
\end{enumerate}

The rest of this paper is organized as follows. Section~\ref{sec: background}
describes the notations and the networked system model. Section~\ref{sec: prob_formulation} elaborates on the problem formulation.
Section~\ref{sec:output_agreement} proposes the generalized PI control that can be implemented fully
decentralizedly and guarantees the convergence of output agreement. Section~\ref{sec:cost_agreement} further proposes the neural-PI control law with local communication, which guarantees both stability criteria
and steady-state optimal resource allocation.
Section~\ref{sec:RL} illustrates how to train the neural-PI control law to optimize the transient performance without jeopardizing stability. Section~\ref{sec:experiments} validates the proposed method through experiments on vehicle platoons and power systems. Section~\ref{sec:conclusion} concludes the paper.


\section{Preliminaries and Background}
\label{sec: background}
\subsection{Notations and preliminaries}
Throughout this manuscript, vectors are denoted in lower-case bold and matrices are denoted in upper-case bold, unless otherwise specified. Vectors of all ones and zeros are denoted as  $\mathbbold{1}_n, \mathbbold{0}_n \in \real^n$, respectively.  Superscript $^*$ indicates the equilibrium value of a variable.
For $\bm{A} \in \mathbb{R}^{m \times n},[\bm{A}]_i$ and $\bm{A}_{i, j}$ represent its $i$-th row and $(i, j)$-th element, respectively. We denote $\mathcal{N}(\bm{A})$ as the null space of matrix $\bm{A}$.  
A continuous function $\bm{g}: \mathcal{D} \mapsto \mathbb{R}^n$ is said to be strictly increasing on $\mathcal{D}\subset  \mathbb{R}^n$ if $(\bm{g}(\bm{\eta})-\bm{g}(\bm{\xi}))^\top(\bm{\eta}-\bm{\xi}) \geq 0$ $\forall\bm{\eta}, \bm{\xi} \in \mathcal{D}$, with equality holds if and only if $\bm{\eta}=\bm{\xi}$. If there further exists $\epsilon>0$ such that  $(\bm{g}(\bm{\eta})-\bm{g}(\bm{\xi}))^\top(\bm{\eta}-\bm{\xi}) \geq \epsilon ||\bm{\eta}-\bm{\xi}||^2$, then $\bm{g}: \mathcal{D} \mapsto \mathbb{R}^n$ is said to be
strongly increasing.

\subsection{Networked system model}\label{subsec: network}
\begin{figure*}[ht]	
	\centering	\includegraphics[width=0.9\textwidth]{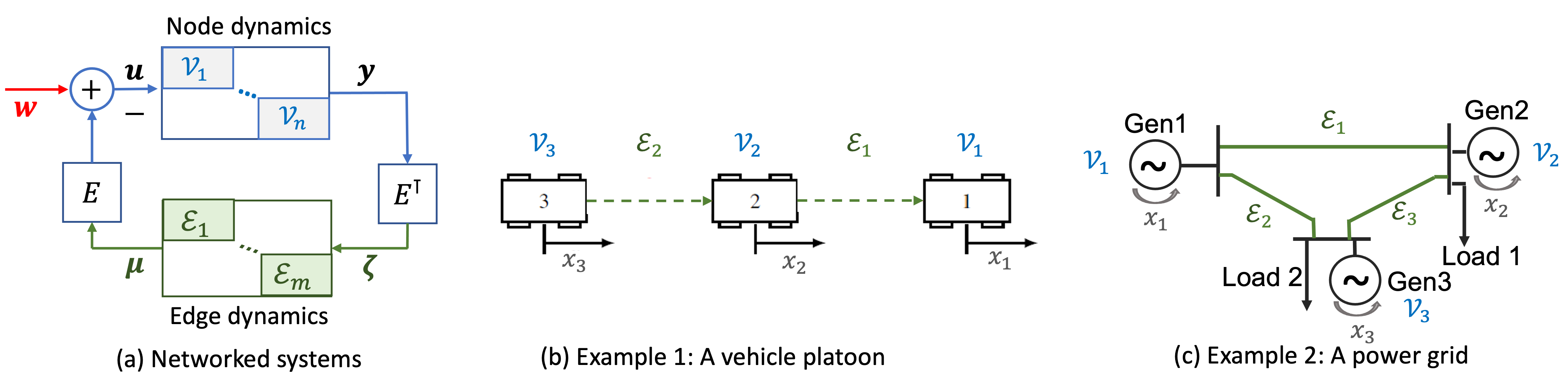}
	 \vspace{-0.2cm}\caption{(a) The networked system with node and edge dynamics, where the inputs and outputs of the nodes and edges are connected by a graph. The external control input is $\bm{w}$. (b) A vehicle platoon where each node is a vehicle   and the edge dynamics describe the relative position between vehicles. (c) A power system where each node is a generator and the edge dynamics are determined by power flow that depends on the angle differences between generators.
 \vspace{-0.4cm}
}
	\label{fig:network}
\end{figure*}

We consider networked systems as illustrated in Fig.~\ref{fig:network}, where the node dynamics (blue blocks) and the edge dynamics (green blocks) form a closed-loop system by coupling their inputs and outputs through a network. Formally, we define the networked dynamical system on an undirected and connected graph $\mathcal{G}=(\mathcal{V}, \mathcal{E})$, consisting of $n$ nodes, $\mathcal{V}:=\left\{v_{1}, \ldots, v_n \right\}$,
and $m$ edges, $\mathcal{E}:=\left\{e_{1}, \ldots, e_m\right\}$. For node $v_i\in \mathcal{V}$ and edge $e_l\in\mathcal{E}$, we will abbreviate them with $i\in \mathcal{V}$ and $l\in\mathcal{E}$.
The incidence matrix
$\bm{E} \in \mathbb{R}^{n \times m}$ is defined such that $\bm{E}_{i,l}$ has value $+1$ if node $i$ is the head of edge $l$, and $-1$ if it is the tail, and $0$ otherwise. For a connected graph, the null space of $\bm{E}^\top$ is $\mathcal{N}(\bm{E}^\top)=\text{range}\left\{\mathbbold{1}_n\right\}$~\cite{biggs1993algebraic}. This is an important property we will use later to show consensus over a networked system.


\subsubsection{Node dynamics} Each node $i\in \mathcal{V}$ represents a single-input single-output (SISO) system, for example, a vehicle in Fig.~\ref{fig:network}(b) or a generator in Fig.~\ref{fig:network}(c). The node dynamics are described by:
\vspace{-0.1cm}
\begin{equation}\label{eq:node_dyn}
\begin{split}
\mathcal{V}_{i}: &
\quad \dot{x}_{i}(t)=f_{i}\left(x_{i}(t), u_{i}(t)\right), \\
& \quad  y_{i}(t)  =h_{i}\left(x_{i}(t)\right), 
\end{split}
\end{equation}
with state $x_{i}(t) \in  \mathbb{R}$, input $u_{i}(t) \in  \mathbb{R}$, and output $y_{i}(t) \in \mathbb{R}$. We assume  functions $f_i$ and $h_i$ are continuously differentiable for all $i$. We sometimes omit the time index $t$ for simplicity. 

\subsubsection{Edge dynamics} For an edge $l\in \mathcal{E}$, its dynamics are defined by 
\vspace{-0.2cm}
\begin{equation}\label{eq:edge_dyn}
\begin{split}
\mathcal{E}_{l}: 
&\quad \dot{\eta}_{l}(t)=\zeta_{l}(t),\\
&\quad \mu_{l}(t)=\psi_{l}\left(\eta_{l}(t)\right),   
\end{split}
\end{equation}
with state $\eta_{l}(t) \in \mathbb{R}$, input $\zeta_{l}(t) \in \mathbb{R}$, and output $\mu_{l}(t) \in \mathbb{R}$. The function $\psi_{l}(\cdot):\mathbb{R}\mapsto\mathbb{R}$ maps the state $\eta_{l}(t)$ of each edge to its output $\mu_{l}(t)$. 


\subsubsection{Interconnection of nodes and edges} In a networked system, the inputs and outputs of nodes and edges are coupled, which forms a closed-loop system shown in Fig.~\ref{fig:network}(a).   For an edge $l$ connecting node $i$ and $j$, its input $\zeta_{l}(t)$ is the difference between the nodal outputs of  node $i$ and $j$, i.e., $\zeta_{l}(t)=y_{i}(t)-y_{j}(t)$. 
The input $u_i$ of a node $i$ consists of the summation of outputs from all connected edges $\sum_{l=1}^m-\bm{E}_{i,l}\mu_l$ and the  external control signal $w_i$.
Because of the lack of real-time communication capacity over the whole network, we consider the  external control signal  following a static control law,  written as  $w_i(y_i):\mathbb{R}\mapsto\mathbb{R}$ where the input is the local observation $y_i$. Consequently, $u_i=-\sum_{l=1}^m\bm{E}_{i,l}\mu_l +w_i(y_i)$. In a vector form, we have the closed-loop system in Fig.~\ref{fig:network}(a) formed by~\eqref{eq:node_dyn},~\eqref{eq:edge_dyn} and their interconnects as follows
\begin{subequations}\label{eq:couple}
    \begin{align}
        \dot{\bm{x}}&=\bm{f}(\bm{x}, \bm{u}),\qquad \bm{y}=\bm{h}(\bm{x})\,,\\      
        \bm{u} &= -\bm{E}\bm{\psi}(\bm{\eta})+ \bm{w}(\bm{y}),\label{eq:control-u}\\ 
        \dot{\bm{\eta}}&=\bm{\zeta},\;\bm{\zeta}=\bm{E}^{\top}\bm{y}, \;\bm{\mu}=\bm{\psi}(\bm{\eta})\,,\label{eq:deta}
    \end{align}
\end{subequations}
where $\bm{x}:=\left(x_i, i \in [n] \right) \in \real^n$, $\bm{y}:=\left(y_i, i \in [n] \right) \in \real^n$, $\bm{u}:=\left(u_i, i \in [n] \right) \in \real^n$, $\bm{\eta}:=\left(\eta_l, l \in [m] \right) \in \real^m$, $\bm{f}:=\left(f_i, i \in [n] \right)$, $\bm{h}:=\left(h_i, i \in [n] \right)$, and $\bm{w}:=\left(w_i, i \in [n] \right)$.

For the networked systems in Fig.~\ref{fig:network}, we wish to achieve two main objectives: 1) fast convergence of system states to the desired operating point; 2) maintaining system operation at the desired operating point with minimal cost. We will elaborate on these two objectives in Section~\ref{sec:stability_criteria} and Section~\ref{sec:optimization_criteria}.
These objectives can be achieved by adjusting the law of external control $\bm{w}(\cdot)$, and if possible, the edge feedback function $\bm{\psi}(\cdot)$. We provide two motivating examples that fall under our model and will be used in experiments in Section~\ref{sec:experiments}. 

\begin{example}[Vehicle platooning] The first example is the vehicle traffic model in Fig.~\ref{fig:network}(b), where each node is a vehicle with velocity $x_i$, and the edge states are the relative position between neighbouring vehicles~\cite{burger2014duality},  i.e., $\dot{\eta}_l = x_i - x_j$ for  neighbouring vehicle $i$ and  $j$. The external control signal $\bm{w}(\cdot)$ sets the nominal velocities for vehicles, and the edge feedback function $\bm{\psi}(\cdot)$ controls how each vehicle responds to the observed differences in velocities. 
The objectives include 1) choose $\bm{w}(\cdot)$ and $\bm{\psi}(\cdot)$ such that all vehicles reach the same velocity; 2) minimize the fuel consumption of vehicles at this velocity.
\end{example}

\begin{example}[Frequency control in power systems] The second example is the power system shown in Fig.~\ref{fig:network}(c), where each node is a generator that rotates with speed (i.e., frequency) $x_i$, and the edge states  are the relative angle difference between them, $\dot{\eta}_l = x_i - x_j$~\cite{cui2022tps} for the line $l$ from generator $i$ and $j$. The external control signal $\bm{w}(\cdot)$ is the adjustment to generator power outputs. Here the edge feedback function $\bm{\psi}(\cdot)$ are determined by physics and are not design variables. The objectives include: 1) choose $\bm{w}(\cdot)$ such that each generator reaches the nominal frequency (e.g., 60Hz); 2) minimize the cost of power generation to maintain this frequency.
\end{example}

\section{Problem Formulation}\label{sec: prob_formulation}
In this paper, we aim to design the external control $\bm{w}$, and if possible, the edge feedback function $\bm{\psi}$ to optimize both the transient and steady-state performances. 
In particular, the prerequisite for the design is that the system should guarantee the stability criteria, i.e., asymptotic stability and output agreement. Below, we first illustrate the stability criteria and then elaborate on the formulation of optimization problems.

\vspace{-0.1cm}
\subsection{Stability criteria}
\label{sec:stability_criteria}

The state of the closed-loop system~\eqref{eq:couple} is $\bm{\varphi}(t):=(\bm{x}(t), \bm{\eta}(t))$, and its equilibrium is $\bm{\varphi}^*:=(\bm{x}^*, \bm{\eta}^*)$.
We adopt the following classical notion  of asymptotic stability around an equilibrium. 

\begin{definition}[Local asymptotic stability~\cite{khalil1996nonlinear}]\label{def: stability}
The networked system~\eqref{eq:couple}
is asymptotically stable around an equilibrium $\bm{\varphi}^*$ if, $\forall \epsilon>0$, $\exists \delta>0$ such that $\|\bm{\varphi}(0) - \bm{\varphi}^*\|<\delta$ ensures $\|\bm{\varphi}(t) - \bm{\varphi}^*\|<\epsilon$, $\forall t\geq0$, and $\exists \delta^\prime>0$ such that $\|\bm{\varphi}(0) - \bm{\varphi}^*\|<\delta^\prime$ ensures $\lim _{t \rightarrow \infty} \|\bm{\varphi}(t) - \bm{\varphi}^*\|=0$. 
\end{definition}
For networked systems described in Fig.~\ref{fig:network}, we are interested in achieving a special type of equilibrium, where the outputs achieve agreement. 
\begin{definition}[Output agreement]\label{def: agreement}
The networked system~\eqref{eq:couple} is said to reach output agreement at $\bm{y}^*=\hat{y}\mathbbold{1}_n$, if $\lim _{t \rightarrow \infty}  y_{i}(t) = \hat{y}$, $\forall i\in\mathcal{V}$, with $\hat{y}\in\real$ being a constant.
\end{definition}



\subsection{Optimization criteria for transient and steady-state performances}
\label{sec:optimization_criteria}

\subsubsection{Transient performance: } The period after a disturbance and before the system settles down to a steady state is called the \emph{transient period}.  During the transient period, our goal is to quickly drive the system to the steady state with the desired agreement value $\Bar{y}$, while minimizing the external control effort $\bm{w}$. 
Thus, for all node $i\in\mathcal{V}$, we consider the cost function $J_i$ that penalizes the output deviation $\left(y_i(t)-\Bar{y}\right)$ and the control effort from $w_i(t)$.
The transient optimization problem up to time $T$ is
\vspace{-0.1cm}
\begin{subequations}\label{eq: transient_optimization}
\begin{align}
    \min_{\bm{w}(\cdot), \bm{\psi}(\cdot)}& \int_{t=0}^T\sum_{i=1}^n J_i(y_i(t)-\Bar{y}, w_i(y_i(t)))dt\,, \label{eq:transient_cost}\\
    \text{s.t. } & 
    \text{dynamics in } \eqref{eq:couple},
     \quad \\
    & \text{stability criteria in Definitions~\ref{def: stability}\text{ and }\ref{def: agreement}},
    \end{align}
\end{subequations}
which is a general formulation including the design freedom in both the external control $\bm{w}(\cdot)$ and the edge feedback function $\bm{\psi}(\cdot)$. For problems where $\bm{\psi}(\cdot)$ is fixed (e.g., frequency control in power systems), we eliminate $\bm{\psi}(\cdot)$ from the optimization variable. 
In practice, 
the system dynamics \eqref{eq:couple} can be highly nonlinear, making it challenging to solve~\eqref{eq: transient_optimization} using conventional optimization techniques. Therefore, the current state-of-the-art is to learn functions $\bm{w}(\cdot)$ and $\bm{\psi}(\cdot)$ by parameterizing them as neural networks and train them by minimizing the cost in~\eqref{eq:transient_cost}.
But the key challenge with applying these neural network-based controllers is guaranteeing stability. 
Even if the learned policy may appear ``stable'' during training, it is not necessarily stable during testing. This can be observed in the vehicle and power system experiments in Section~\ref{sec:experiments}. 

\subsubsection{Optimal resource allocation at the steady-state} In addition to optimizing the transient period performance, we also want to optimize the steady-state cost, i.e. the cost of maintaining the system outputs at the desired value $\Bar{y}\mathbbold{1}_n$. 
For example, the frequencies in a power system (in North America) should be very close to 60 Hz~\cite{sauer2017power}. Since there are many ways to set steady-state generator power outputs to achieve this, the system operator needs to find the one that minimizes the generation cost. Let $C_i(\cdot):\real\mapsto\real$ be the cost function for the external control of node $i\in\mathcal{V}$. The optimal resource allocation problem is
\vspace{-0.2cm}
\begin{equation}\label{eq: allocation_optimization}
\begin{split}
&\min_{\bm{w}^*}C(\bm{w}^*):=\sum_{i=1}^nC_i(w_i^*),\\ 
&\quad \text{s.t. }
\bm{y}^*=\Bar{y}\mathbbold{1}_n.    
\end{split}
\end{equation}
The goal is to enforce that the external control $\bm{w}^*$ at the steady state solves~\eqref{eq: allocation_optimization}, which indicates that the setpoints of the control action settle down to the optimal resource allocation solution. 





\vspace{-0.1cm}

\subsection{Bridging controller design and stability via passivity analysis}

The transient optimization \eqref{eq: transient_optimization} and steady-state optimization \eqref{eq: allocation_optimization} appear to be two different optimization problems.
Thus, the key question we address in this paper is:   how to find a learning-based controller that solves \eqref{eq: transient_optimization} and \eqref{eq: allocation_optimization} simultaneously, while guaranteeing system stability?

To bridge controller design and stability of the networked systems, we use the property of passivity. Passivity is a widely adopted tool
to analyze stability in control of networked systems~\cite{slotine1991applied}. 
Specifically, we consider the notion of equilibrium-independent passivity (EIP)~\cite{arcak2016networks,hines2011equilibrium} defined below.
\begin{definition}[Equilibrium-independent passivity~\cite{arcak2016networks}]\label{def: EIP}
 The system described by
$\dot{x_i}=f_i(x_i, u_i),
 y_i=h_i(x_i), x_i\in \real , u_i\in  \real$
 is  equilibrium-independent
passive (EIP) if  there exists a nonempty set $ \mathcal{U}_i^*\subseteq \real$ such that 
for every $u_i^*\in \mathcal{U}_i^*$, (i)  there exists a unique $x_i^*$ satisfying $f_i(x_i^*, u_i^*)=0$, and  (ii) there exists
a positive definite storage function
$W_{i}\left(x_{i}, x_{i}^*\right)$ such that, $\forall x_i\in \real, u_i\in \real$,
 \begin{equation}\label{eq: def_EIP}
W_{i}\left(x_{i}^*, x_{i}^*\right)=0 \ \mbox{and}\ \dot{W}_{i}\left(x_{i}, x_{i}^*\right) \leq\left(y_{i}-y_{i}^*\right)\left(u_{i}-u_{i}^*\right).     
 \end{equation}
If there further exists  a constant $\rho_{i}>0$ such that
\begin{equation}\label{eq: def_strict_EIP}
\dot{W}_{i}\left(x_{i}, x_{i}^*\right) \leq-\rho_{i}\left\|y_{i}-y_{i}^*\right\|^{2}+\left(y_{i}-y_{i}^*\right)\left(u_{i}-u_{i}^*\right)\,,  
\end{equation}
then the system is strictly EIP.
\end{definition}
\begin{remark}[Comparison between passivity and EIP]
The conventional definition of passivity is referenced to a
chosen equilibrium input-state-output configuration,
which is typically taken to be the origin~\cite{slotine1991applied, simpson2018equilibrium}. However, the equilibrium is obtained by the simultaneous solution of all subsystems and is sometimes difficult to be 
explicitly computed.
The notion of EIP characterizes
dynamical systems that are passive with respect to an arbitrary
equilibrium point, which enables
a convergence or stability analysis of  networked systems without computing
the equilibrium a prior~\cite{arcak2016networks,hines2011equilibrium}.
\end{remark}

We make the following assumption that each node is strictly EIP, which  as an abstraction of the system would allow us to design generalized controllers for networked systems.

\begin{assumption}[Strictly EIP of nodes]\label{ass: EIP_node}
$\forall i\in\mathcal{V}$, the node dynamics~\eqref{eq:node_dyn} is strictly EIP
with a storage function $W_{i}^\mathcal{V}\left(x_{i}, x_{i}^*\right)$.
\end{assumption}

The EIP property in Assumption~\ref{ass: EIP_node} has been found in a large class of physical systems, including transportation~\cite{burger2014duality}, power systems~\cite{cui2022equilibrium,nahata2020passivity}, robotics~\cite{meissen2017passivity}, communication~\cite{simpson2018equilibrium}, and others. For instance, for power system frequency control, a quadratic function $W_{i}^\mathcal{V} = \frac{1}{2}(x_i - x_i^*)^2$ can serve as the storage function for each node and verify the EIP assumption.

Note that an arbitrary interconnection of nodes that satisfy the EIP property does not necessarily lead to a stable system, and it is important to \emph{design} the interconnections and controller to satisfy certain conditions to achieve stability. The rest of the paper characterizes algebraic conditions that the controllers need to satisfy, and how neural networks can be structured to satisfy these conditions.

\section{Neural-PI Control with Stability and Output Agreement Guarantees}
\label{sec:output_agreement}
In this section,
we propose a generalized proportional-integral (PI) structure for external control $\bm{w}$ and conditions on the edge feedback function $\bm{\psi}$, for network systems without communication capability. In particular, we show that monotonicity of $\bm{w}$ and $\bm{\psi}$ implies the convergence of the output agreement to the required value. 

\subsection{ Generalized PI control with stability implied by monotonicity}
We start by writing the external controller $\bm{w}$ as the following form, for each node $i\in\mathcal{V}$, 
\begin{subequations}\label{eq:general_PIcontroller}
\begin{align}
    w_i &= p_{i}(\bar{y}_i - y_i) + r_i(s_i)\, ,\\
    \dot{s}_i & = \bar{y}_i - y_i\, .
\end{align}
\end{subequations}
At each node $i$, the controller is made up of two components. The component $p_i(\cdot)$  is a proportional term, which is a function of the tracking error between the current output $y_i$ and desired output value $\bar{y}_i$. The component $r_i(\cdot)$ is the integral term, which is a function of the integral of historical tracking errors denoted by $s_i$ where $\dot{s}_i = \Bar{y}_i-y_i $, $s_i(0)=0$. The above controller follows the structure of the widely adopted PI controller~\cite{slotine1991applied, khalil1996nonlinear}. Intuitively, the proportional term drives $y_i$ close to $\Bar{y}_i$ and the integral term drives the accumulated tracking error to zero. 


In most existing applications, linear PI controllers are used~\cite{Zhao2015acc, andreasson2014distributed,  weitenberg2018exponential, golnaraghi2017automatic}, where $p_{i}(\bar{y}_i - y_i) = K_{i,1}(\bar{y}_i - y_i)$ and $r_i(s_i) = K_{i,2} s_i$ with $K_{i,1}\in\real$ and $K_{i,2}\in\real$ being constants. Linear PI controllers are easy for  implementation and analysis. 
However, their transient performance can be poor for large-scale nonlinear systems.

Here, we consider a generalized PI controller by letting $p_{i}(\cdot)$ and $r_{i}(\cdot)$ in \eqref{eq:general_PIcontroller} be nonlinear functions parameterized by neural networks. 
Training the neural networks may find parameterizations of controllers that attain lower transient cost in~\eqref{eq: transient_optimization} than linear PI control. However, it is challenging to guarantee the stability criteria in Section~\ref{sec:stability_criteria} for controllers parameterized by general neural networks. We seek to overcome this challenge through structured design: we first derive the structure of controllers that attain the required stability and output agreement guarantees, then design neural networks to satisfy the structure. 
 The key structure we use are monotonically increasing functions, and the structured controller design is given in Design \ref{design: agreement_set}. We will prove that this controller design guarantees the convergence of output to the desired agreement in the next subsection. 

\begin{design}[Output agreement guarantees]\label{design: agreement_set}
 The external control $w_i$ for each node $i\in\mathcal{V}$ is given by \eqref{eq:general_PIcontroller}, where the functions $p_i(\cdot):\mathbb{R}\mapsto\mathbb{R}$ and $r_i(\cdot):\mathbb{R}\mapsto\mathbb{R}$ are Lipschitz continuous and strongly increasing with $p_i(0)=0$, $r_i(0)=0$. In addition, 
 the functions $\psi_{l}(\cdot)$ are strongly increasing for each edge $l\in\mathcal{E}$. Compactly, the designed controller \eqref{eq:control-u} is
\begin{subequations}\label{eq: PIcontroller1}
\begin{align}
    \bm{u} &=-\bm{E}\bm{\psi}(\bm{\eta}) + \bm{p}(-\bm{y}+\Bar{y}\mathbbold{1}_n) +  \bm{r}(\bm{s})\,,\label{subeq: PIcontroller1_u}\\
    \dot{\bm{s}} &= -(\bm{y}-\Bar{y}\mathbbold{1}_n)\,,\label{subeq: s_bary}
\end{align}    
\end{subequations}
where $\bm{s}:=\left(s_i, i \in [n] \right) \in \real^n$, $\bm{p}:=\left(p_i, i \in [n] \right)$, and $\bm{r}:=\left(r_i, i \in [n] \right)$.
\end{design}

The above controller design features two parts: the external controller $\bm{w}$ (equals to $\bm{p}(-\bm{y}+\Bar{y}\mathbbold{1}_n) +  \bm{r}(\bm{s})$), and if possible, the edge dynamics $\bm{\psi}(\cdot)$. The external controller $\bm{w}$ generalizes the linear PI controller, allowing both $p_i(\cdot)$ and $r_i(\cdot)$ to be nonlinear functions, as long as they are strongly increasing and cross the origin 
 (these
 are sometimes called 
 class $\mathcal{K}$ functions in the literature~\cite{arcak2002input}).

For systems where we have the design freedom on $\psi_l(\eta_l)$  (e.g., vehicle platooning in Fig.~\ref{fig:network} (b)), Controller Design~\ref{design: agreement_set} provides the algebraic constraint on the function $\bm{\psi}(\cdot)$. Although this condition on $\bm{\psi}(\cdot)$ is also presented in~\cite{coogan2014dissipativity,burger2014duality}, they did not consider how to design a good $\bm{\psi}(\cdot)$. In fact, \cite{coogan2014dissipativity} and~\cite{burger2014duality}  choose two very different functions,  $\psi_l(\eta_l) = \tanh(\eta_l)$ and $\psi_l(\eta_l) = (\eta_l)^{1/3}$, respectively. As we will show in our experiments, neither is close to being optimal for the transient performance. To search for potentially better functions, Section~\ref{sec:increasing_NN} will elaborate on how to parameterize neural networks such that these monotonicity conditions in Controller Design~\ref{design: agreement_set} can be met, and how the controllers and edge dynamics can be optimized through training.

  
\begin{remark}[Extended edge feedback function]
Sometimes $\psi_{l}(\eta_l)$ is only defined for the region  $\eta_l\in[\underline{\eta}_l, \overline{\eta}_l]$. In this case, we extend the definition such that 
\begin{equation*}
    \psi_{l}(\eta_l)=\left\{\begin{matrix}
  \psi_{l}(\underline{\eta}_l)+L\left(\eta_l-\underline{\eta}_l\right) &  \eta_l<\underline{\eta}_l\\
 \psi_{l}(\eta_l) &  \eta_l\in[\underline{\eta}_l, \overline{\eta}_l]\\
  \psi_{l}(\overline{\eta}_l)+L\left(\eta_l-\overline{\eta}_l\right) &  \eta_l>\overline{\eta}_l\\
\end{matrix}\right. \, ,
\end{equation*}
where $L>0$ is the Lipschtiz constant of $\psi_{l}(\eta_l)$. Note that we slightly overload the notation $\psi_{l}(\eta_l)$ to simplify the notation. This function is well-defined on $\real$ and will be used later to construct a radially unbounded Lyapunov function. The extended definition is only utilized for deriving quadratic bounds for Lyapunov functions, and we do not need to specify $L$ when optimizing over $\psi_l(\cdot)$ in the region $[\underline{\eta}_l, \overline{\eta}_l]$. 
\end{remark}

\subsection{Convergence to the desired output agreement}

The next  theorem shows that the output   of the system under Controller Design~\ref{design: agreement_set} converges to the desired agreement level.

\begin{theorem}[Convergence of output agreement]~\label{thm: output_level_stable}
Suppose that Assumption~\ref{ass: EIP_node}
holds and the input $\bm{u}$ follows~\eqref{eq: PIcontroller1} in Controller Design~\ref{design: agreement_set}.
Suppose the storage function $W_{i}^\mathcal{V}\left(x_{i}, x_{i}^*\right)$ is radially unbounded, $\forall i\in\mathcal{V}$. 
If  the system~\eqref{eq:couple} has a feasible equilibrium, 
then the output of each node  converges to the desired steady-state value $\bar{y}$, i.e., $\bm{y}^*=\Bar{y}\mathbbold{1}_n$.

\end{theorem}

By Theorem~\ref{thm: output_level_stable}, the Controller Design~\ref{design: agreement_set} provides key structural property for the controllers to have provable guarantees on convergence to the desired output agreement.  We show the convergence of the output by constructing a positive definite function $V_1(\bm{x},\bm{\eta},\bm{s} )|_{\bm{x}^*,\bm{\eta}^*,\bm{s}^*}$ that is  radially unbounded and  the time derivative $\dot{V}_1(\bm{x},\bm{\eta},\bm{s} )|_{\bm{x}^*,\bm{\eta}^*,\bm{s}^*}\leq -\sum_{i=1}^{n} \rho_{i}\left\|y_{i}-y_{i}^*\right\|^{2}$. Then the convergence of $\sum_{i=1}^{n} \rho_{i}\left\|y_{i}-y_{i}^*\right\|^{2}$ to zero follows directly from Barbalat's lemma~\cite[Lemma~4.2]{slotine1991applied}. The rest of this section outlines the proof of Theorem~\ref{thm: output_level_stable}.

\begin{remark}[Robustness]
Theorem~\ref{thm: output_level_stable} does not depend on the specifics of $f_i(\cdot)$ in nodal dynamics \eqref{eq:node_dyn} (as long as it is EIP), making the convergence certification robust to parameter changes for systems satisfying Assumption~\ref{ass: EIP_node}. This will be demonstrated in the experiment on power system control.
\end{remark}

\begin{remark}[Equilibrium]\label{rmk:Equilibrium}
Observe from \eqref{eq:couple} and \eqref{subeq: s_bary} that the equilibrium is given by $\dot{\bm{x}}=\mathbbold{0}_n$, $\dot{\bm{\eta}}=\mathbbold{0}_m$, and $\dot{\bm{s}}=\mathbbold{0}_n$, which yields $\bm{f}(\bm{x}^*,\bm{u}^*)=\mathbbold{0}_n$, $\bm{\zeta}^*=\bm{E}^{\top}\bm{y}^*=\mathbbold{0}_m$, and $\bm{y}^*=\Bar{y}\mathbbold{1}_n$. Thus,
the set of feasible equilibrium of system~\eqref{eq:couple} under Controller Design~\ref{design: agreement_set} is
$\mathcal{S}_e:=\big\{\bm{x}^*,\bm{\eta}^*,\bm{s}^*|
\bm{f}(\bm{x}^*,\bm{u}^*)=\mathbbold{0}_n, \Bar{y}\mathbbold{1}_n=\bm{h}(\bm{x}^*), \bm{r}\left(\bm{s}^*\right)-\bm{E}\bm{\psi}(\bm{\eta}^*)\allowbreak=\bm{u}^*,  \bm{u}^*\in\mathcal{U}^*   \big\}$. Without extra assumption, there might be multiple feasible equilibria in $\mathcal{S}_e$.
This is the reason why  we show the convergence of the output instead of the stability to a unique equilibrium.
In Section~\ref{sec:cost_agreement},  we will further show the controller design that yields a unique equilibrium by enforcing that  $\bm{r}\left(\bm{s}^*\right)$ settles down to the solution of optimal resource allocation. 
\end{remark}




 To show the convergence of the output, we construct a positive definite function $V_1(\bm{x},\bm{\eta},\bm{s} )|_{\bm{x}^*,\bm{\eta}^*,\bm{s}^*}$  using the storage functions $W_{i}^\mathcal{V}(x_i,x_i^*), i\in\mathcal{V}$ as well as the integral functions 
\begin{equation}\label{eq:integ}
 R(\bm{s}):=\sum_{i=1}^n\! \int_0^{s_i}\!\!r_i(z) \mathrm{d}z\ \ \text{and}\ \ Q(\bm{\eta}):=\sum_{l=1}^m\! \int_0^{\eta_{l}}\!\!\psi_{l}(z) \mathrm{d}z\end{equation}
associated with  the monotone functions $r_i(\cdot)$ and $\psi_{l}(\cdot)$ in Controller Design~\ref{design: agreement_set}. Namely, we construct a function 
 \begin{align}\label{eq:Lyapunov_function}
    &V_1(\bm{x},\bm{\eta},\bm{s} )|_{\bm{x}^*,\bm{\eta}^*,\bm{s}^*}\nonumber\\ 
    &:= \sum_{i=1}^{n} W_{i}^\mathcal{V}(x_i,x_i^*) +B^\mathcal{V}(\bm{s}, \bm{s}^*) +B^\mathcal{E}(\bm{\eta},\bm{\eta}^*)\,,   
\end{align}
where 
$B^\mathcal{V}(\bm{s}, \bm{s}^*)$ and $B^\mathcal{E}(\bm{\eta},\bm{\eta}^*)$ are 
 the Bregman distances associated with the integral functions $R(\bm{s})$ and $Q(\bm{\eta})$, i.e., 
\begin{subequations}\label{eq:Bregman_s}
\begin{align}
    B^\mathcal{V}\!\left(\bm{s}, \bm{s}^*\right)\!& :=\! R\left(\bm{s}\right)-R\left(\bm{s^*}\right)-\!\nabla R (\bm{s^*})^\top\left(\bm{s}-\bm{s^*}\right)\,,\label{eq:Bv}\\
    B^\mathcal{E}\!\left(\bm{\eta}, \bm{\eta}^*\right)\!&:=\!
    Q\left(\bm{\eta}\right)-Q\left(\bm{\eta}^*\right)-\!\nabla Q\left(\bm{\eta}^*\right)^\top\!\left(\bm{\eta}-\bm{\eta}^*\right)\,.
\end{align}
\end{subequations}
The Bregman distances $B^\mathcal{V}(\bm{s}, \bm{s}^*)$ and $B^\mathcal{E}(\bm{\eta},\bm{\eta}^*)$ are lower bounded by quadratic forms due to the following lemma. 

\begin{lemma}[Bregman distances of monotone functions]\label{lem:Bregman}
For $r_i(\cdot):\real\mapsto\real$ and $\psi_{l}(\cdot):\real\mapsto\real$ that are Lipschitz continuous and strongly increasing, 
there exist some $\epsilon_\mathrm{v}>0$ and $\epsilon_\mathrm{e}>0$ such that the Bregman distances 
in~\eqref{eq:Bregman_s} satisfy
\begin{equation}\label{eq:B-quad}
B^\mathcal{V}(\bm{s}, \bm{s}^*)\geq \frac{\epsilon_\mathrm{v}}{2}||\bm{s}-\bm{s}^*||_2^2\,,    B^\mathcal{E}(\bm{\eta}, \bm{\eta}^*)\geq \frac{\epsilon_\mathrm{e}}{2}||\bm{\eta}-\bm{\eta}^*||_2^2\,.
\end{equation}
\end{lemma}
\begin{proof}
We begin by showing that $R(\bm{s})$ defined in \eqref{eq:integ} is strongly convex.
Since $r_i(\cdot)$ is strongly increasing, there exists $\epsilon_i>0$ such that 
\begin{equation}\label{eq:mono-r}
\left(r_i(s_i)-r_i(s_i^\prime)\right)\left(s_i-s_i^\prime\right)\geq\epsilon_i\left(s_i-s_i^\prime\right)^2, \forall s_i, s_i^\prime \in \real\,.    
\end{equation}
Then, note that, $\forall\bm{s}\neq\bm{s}^\prime$,
\begin{equation}\label{eq:mono-gradL}
    \begin{split}
    &\left(\nabla R (\bm{s})-\nabla R (\bm{s}^\prime)\right)^\top\left(\bm{s}-\bm{s}^\prime\right)\\
    =&\left(\bm{r}(\bm{s})-\bm{r}(\bm{s}^\prime)\right)^\top\left(\bm{s}-\bm{s}^\prime\right)\\
    =&\sum_{i=1}^n\left(r_i(s_i)-r_i(s_i^\prime)\right)\left(s_i-s_i^\prime\right)\\\geq&\sum_{i=1}^n\epsilon_i\left(s_i-s_i^\prime\right)^2    \geq\underbrace{\min_{i\in[n]}\epsilon_i}_{:=\epsilon_\mathrm{v}}\|\bm{s}-\bm{s}^\prime\|_2^2\,,        
    \end{split}
\end{equation}
where the first inequality results from \eqref{eq:mono-r}. By~\cite[Chapter IV, Theorem~4.1.4]{hiriart1996convex}, \eqref{eq:mono-gradL} indicates that $R(\bm{s})$ is $\epsilon_\mathrm{v}$-strongly convex, which further implies that $B^\mathcal{V}(\bm{s}, \bm{s}^*)$ defined in~\eqref{eq:Bv} satisfies \eqref{eq:B-quad} by~\cite[Chapter IV, Theorem~4.1.1]{hiriart1996convex}.
The result for $B^\mathcal{E}(\bm{\eta},\bm{\eta}^*)$ follows from the same line of argument.
\end{proof}

With Lemma~\ref{lem:Bregman}, it is straightforward to show that $V_1(\bm{x},\bm{\eta},\bm{s} )|_{\bm{x}^*,\bm{\eta}^*,\bm{s}^*}$ is positive definite and radially unbouded.

\begin{lemma}[Positive definiteness of $V_1(\bm{x},\bm{\eta},\bm{s} )|_{\bm{x}^*,\bm{\eta}^*,\bm{s}^*}$]\label{lem:well_definied_Lyap}
Suppose assumptions in Theorem~\ref{thm: output_level_stable} hold, then $V_1(\bm{x},\bm{\eta},\bm{s} )|_{\bm{x}^*,\bm{\eta}^*,\bm{s}^*}$  is a radially unbounded 
function satisfying
$V_1(\bm{x}^*,\bm{\eta}^*,\bm{s}^* )|_{\bm{x}^*,\bm{\eta}^*,\bm{s}^*}=0$  and $V_1(\bm{x},\bm{\eta},\bm{s} )|_{\bm{x}^*,\bm{\eta}^*,\bm{s}^*}>0$, $\forall (\bm{x},\bm{\eta},\bm{s} )\neq(\bm{x}^*,\bm{\eta}^*,\bm{s}^* )$.
\end{lemma}
\begin{proof} 
 By Lemma~\ref{lem:Bregman}, $B^\mathcal{V}(\bm{s}, \bm{s}^*)$ and $B^\mathcal{E}(\bm{\eta},\bm{\eta}^*)$ are radially unbounded for $\bm{s}\in\real^n$ and $\bm{\eta}\in\real^m$, respectively. By assumption in Theorem~\ref{thm: output_level_stable}, $\sum_{i=1}^{n} W_{i}^\mathcal{V}(x_i,x_i^*) $ is radially unbounded for $\bm{x}\in\real^n$. Hence, $V_1(\bm{x},\bm{\eta},\bm{s} )|_{\bm{x}^*,\bm{\eta}^*,\bm{s}^*}$  defined by~\eqref{eq:Lyapunov_function} is radially unbounded.

We now show that $V_1(\bm{x},\bm{\eta},\bm{s} )|_{\bm{x}^*,\bm{\eta}^*,\bm{s}^*}$ is positive definite. By Assumption~\ref{ass: EIP_node}, $\sum_{i=1}^{n} W_{i}^\mathcal{V}(x_i,x_i^*) \geq 0$ with equality holds only when $\bm{x}=\bm{x}^*$. It can been seen clearly from \eqref{eq:B-quad} in Lemma~\ref{lem:Bregman} that $B^\mathcal{E}(\bm{\eta},\bm{\eta}^*)\geq 0 $ and 
$B^\mathcal{V}(\bm{s}, \bm{s}^*)\geq 0$ with equality holds only when $\bm{\eta}=\bm{\eta}^*$ and $\bm{s}=\bm{s}^*$, respectively. Hence,  $V_1(\bm{x},\bm{\eta},\bm{s} )|_{\bm{x}^*,\bm{\eta}^*,\bm{s}^*}$ is positive definite and equals to zero only at $(\bm{x}^*,\bm{\eta}^*,\bm{s}^*)$.
\end{proof}

We complete the proof of convergence to output agreement by showing the following lemma.
\begin{lemma}[Time derivative of $V_1(\bm{x},\bm{\eta},\bm{s} )|_{\bm{x}^*,\bm{\eta}^*,\bm{s}^*}$]\label{lem:dV}
Let assumptions in Theorem~\ref{thm: output_level_stable} hold. The time derivative of~\eqref{eq:Lyapunov_function} satisfies
$\dot{V}_1(\bm{x},\bm{\eta},\bm{s} )|_{\bm{x}^*,\bm{\eta}^*,\bm{s}^*}\leq -\sum_{i=1}^{n} \rho_{i}\left\|y_{i}-\Bar{y}\right\|^{2}$. Moreover, $\sum_{i=1}^{n} \rho_{i}\left\|y_{i}-\Bar{y}\right\|^{2}\to 0$ as $t\to\infty$.
\end{lemma}
\begin{proof}


To prepare for the calculation of the time derivative of~\eqref{eq:Lyapunov_function}, 
we start by calculating the time derivatives of functions $B^\mathcal{V}(\bm{s}, \bm{s}^*)$ and $B^\mathcal{E}(\bm{\eta},\bm{\eta}^*)$ in \eqref{eq:Bregman_s}. Clearly, the time derivative of $B^\mathcal{V}(\bm{s}, \bm{s}^*)$ is
\vspace{-0.1cm}
\begin{align}\label{eq:dot_bregman_s}
    \dot{B}^\mathcal{V}(\bm{s}, \bm{s}^*) &=\left(\nabla R\left(\bm{s}\right)-\nabla R\left(\bm{s}^*\right)\right)^\top \dot{\bm{s}}\nonumber
    \\
    & \stackrel{\circled{1}}{=}-\left(\bm{r}(\bm{s})-\bm{r}\left(\bm{s}^*\right)\right)^\top(\bm{y}-\Bar{y}\mathbbold{1}_n)\,,
\end{align}
where $\circled{1}$ follows from $\nabla R\left(\bm{s}\right)=\bm{r}\left(\bm{s}\right)$ by \eqref{eq:integ} and the dynamics of $\bm{s}$ in \eqref{subeq: s_bary}. 
Similarly, the time derivative of $B^\mathcal{E}(\bm{\eta},\bm{\eta}^*)$ is
\vspace{-0.1cm}
\begin{align}\label{eq:dot_bregman_eta}
    \dot{B}^\mathcal{E}(\bm{\eta},\bm{\eta}^*) &=\left(\nabla Q\left(\bm{\eta}\right)-\nabla Q\left(\bm{\eta}^*\right)\right)^\top \dot{\bm{\eta}}
    \nonumber\\
    & \stackrel{\circled{1}}{=}\left(\bm{\psi}\left(\bm{\eta}\right)-\bm{\psi}\left(\bm{\eta}^*\right)\right)^\top \bm{\zeta}\,,   
\end{align}
where $\circled{1}$ uses $\nabla Q\left(\bm{\eta}\right)=\bm{\psi}\left(\bm{\eta}\right)$ by \eqref{eq:integ} and $\dot{\bm{\eta}}=\bm{\zeta}$ by \eqref{eq:deta}.

Now, we can compute the time derivative of~\eqref{eq:Lyapunov_function} as
\vspace{-0.1cm}
\begin{align}\label{eq:dotLyap1}
    &\dot{V}_1(\bm{x},\bm{\eta},\bm{s})|_{\bm{x}^*,\bm{\eta}^*,\bm{s}^*}  
    \nonumber\\
    =& \sum_{i=1}^{n} \dot{W}_{i}^\mathcal{V}\left(x_{i}, x_{i}^*\right) +\dot{B}^\mathcal{V}(\bm{s}, \bm{s}^*)+ \dot{B}^\mathcal{E}(\bm{\eta},\bm{\eta}^*)
    \nonumber\\
     \stackrel{\circled{1}}{\leq} &
    -\sum_{i=1}^{n}\rho_{i}\left\|y_{i}-y_{i}^*\right\|^{2}
    +(\bm{y}-\bm{y}^*)^{\top}(\bm{u}-\bm{u}^*)\nonumber\\
    &-\left(\bm{r}(\bm{s})\!-\!\bm{r}(\bm{s}^*)\right)^\top(\bm{y}\!-\!\Bar{y}\mathbbold{1}_n)+\left(\bm{\psi}(\bm{\eta})\!-\!\bm{\psi}(\bm{\eta}^*)\right)^\top \bm{\zeta}
 \nonumber\\    \stackrel{\circled{2}}{=} &
    -\sum_{i=1}^{n}\rho_{i}\left\|y_{i}-\Bar{y}\right\|^{2}-(\bm{y}-\Bar{y}\mathbbold{1}_n)^{\top}\bm{E}\left(\bm{\psi}(\bm{\eta})-\bm{\psi}(\bm{\eta}^*)\right)\nonumber\\&
    +(\bm{y}-\Bar{y}\mathbbold{1}_n)^{\top} \bm{p}(-\bm{y}+\Bar{y}\mathbbold{1}_n) +\bm{\zeta}^\top\left(\bm{\psi}(\bm{\eta})\!-\!\bm{\psi}(\bm{\eta}^*)\right) 
 \nonumber\\    
    \stackrel{\circled{3}}{=}&-\sum_{i=1}^{n} \rho_{i}\left\|y_{i}-\Bar{y}\right\|^{2}- \sum_{i=1}^{n}p_i(\Bar{y}-y_i)(\Bar{y}-y_i)\nonumber\\
    \stackrel{\circled{4}}{\leq} &-\!\sum_{i=1}^{n} \rho_{i}\left\|y_{i}-\Bar{y}\right\|^{2}\!-\!\sum_{i=1}^{n}\tilde{\epsilon}_i\left(\Bar{y}-y_i\right)^2\nonumber\\\leq&-\!\sum_{i=1}^{n} \rho_{i}\left\|y_{i}-\Bar{y}\right\|^{2}\leq0\quad\text{with}\quad\rho_i >0\,.
\end{align}
Here, some tricks are used for constructing a quadratic format. In $\circled{1}$, the strictly EIP property of nodes under Assumption~\ref{ass: EIP_node} and results in \eqref{eq:dot_bregman_s} and \eqref{eq:dot_bregman_eta} are used. In $\circled{2}$, Controller Design~\ref{design: agreement_set} described by \eqref{eq: PIcontroller1} and the corresponding equilibrium property $\bm{y}^*=\Bar{y}\mathbbold{1}_n$ in Remark~\ref{rmk:Equilibrium} are used. 
In $\circled{3}$, $\bm{y}^{\top}\bm{E}=\bm{\zeta}^{\top}$ by \eqref{eq:deta} and $(\Bar{y}\mathbbold{1}_n)^{\top}\bm{E}={\bm{y}^*}^{\top}\bm{E}=\mathbbold{0}_m^{\top}$ by Remark~\ref{rmk:Equilibrium} are used.
In $\circled{4}$, the fact that $p_i(\cdot)$ is strongly increasing with $p_i(0)=0$ is used such that $p_i(\Bar{y}-y_i)(\Bar{y}-y_i)=\left(p_i(\Bar{y}-y_i)-p_i(0)\right)(\Bar{y}-y_i-0)\geq\tilde{\epsilon}_i\left(\Bar{y}-y_i\right)^2$ for some $\tilde{\epsilon}_i>0$, $\forall i\in[n]$.

Recall from Lemma~\ref{lem:well_definied_Lyap} that $V_1(\bm{x},\bm{\eta},\bm{s} )|_{\bm{x}^*,\bm{\eta}^*,\bm{s}^*}$ is radially unbounded and positive definite. Thus, by invoking Barbalat's lemma~\cite[Lemma~4.2]{slotine1991applied}, one can conclude from \eqref{eq:dotLyap1} that $\sum_{i=1}^{n} \rho_{i}\left\|y_{i}-\Bar{y}\right\|^{2}\to 0$ as $t\to\infty$.
\end{proof}
Lemma~\ref{lem:dV} indicates that, if the system~\eqref{eq:couple} under Controller Design~\ref{design: agreement_set} described by \eqref{eq: PIcontroller1} has a feasible equilibrium, 
then the output of each node  converges to a constant steady-state value $\bar{y}$ as $t\to\infty$, i.e., $\bm{y}^*=\Bar{y}\mathbbold{1}_n$. This concludes the proof of output agreement in Theorem~\ref{thm: output_level_stable}.

\section{Neural-PI Control for Distributed Optimal Resource Allocation}
\label{sec:cost_agreement}
In this section, we extend the controller design in Section~\ref{sec:output_agreement} to realize the optimal steady-state resource allocation. Unlike the controller in~\eqref{eq: PIcontroller1} that only relies on local information,  the  optimization of steady-state cost requires communication between neighbours. This section first reformulates the steady-state problem~\eqref{eq: allocation_optimization} and derives its optimality conditions, then presents a distributed algorithm that is embedded into the controller design to meet these conditions.

\subsection{Optimal resource allocation at steady-state}
\label{sec:optimal_resource_allocation}
Ensuring that the steady-state actions settle down to the solution of the  resource allocation problem~\eqref{eq: allocation_optimization} is in general challenging, since the steady-state optimization cannot be trivially incorporated into the controller and the transient optimization~\eqref{eq: transient_optimization}. To overcome this challenge, we contribute to deriving an equivalent formulation for the constraint $\bm{y}^*=\Bar{y}\mathbbold{1}_n$, which subsequently converts the optimality condition of~\eqref{eq: allocation_optimization} into a consensus condition that can be incorporated into the controller design in~\eqref{eq: PIcontroller1}. 

To derive this condition, we first establish an assumption that is basic for the stability analysis of networked systems where each node is a SISO system represented by~\eqref{eq:node_dyn}.
\begin{assumption}[Equilibrium input-state-output mapping of node dynamics]\label{ass:mapping}$\forall i\in\mathcal{V}$,
there exists a  continuous function 
 $k_{x,i}(\cdot): \mathcal{U}_i^* \mapsto \real$
for the equilibrium input-state map such that $x_i^{*}=k_{x,i}(u_i^{*})$ and $f_i\left(k_{x,i}\left(u_i^{*}\right), u_i^{*}\right)=0$. Moreover, $h_i(\cdot)$ and $k_{x,i}(\cdot)$ are bijective functions with inverse functions given by $h_i^{-1}(\cdot)$ and $k_{x,i}^{-1}(\cdot)$, respectively, such that  the equilibrium input-output map
$\left(h_i\circ k_{x,i}\right)(\cdot): \mathcal{U}_i^* \mapsto \real$, i.e., $y_i^{*}=h_i\left(k_{x,i}\left(u_i^{*}\right)\right)$ is a bijection as well.
\end{assumption}

Assumption \ref{ass:mapping} is required to show the uniqueness of equilibrium and derive the optimality conditions for the resource allocation problem~\eqref{eq: allocation_optimization}.
It can be easily checked given $f_i(\cdot)$.

The following lemma presents an equivalent formulation for  the steady-state resource allocation problem and gives its optimality conditions. 

\begin{lemma}[Equivalent formulation for optimal resource allocation]\label{lem: Equivalent_opf}
Let Assumption~\ref{ass:mapping} hold and suppose that 
the optimal resource allocation problem~\eqref{eq: allocation_optimization} has a feasible solution at the output agreement level $\Bar{y}$. Then the optimization problem~\eqref{eq: allocation_optimization}  is equivalent to
\vspace{-0.3cm}
\begin{subequations}\label{eq: optimization_equivalent}
\begin{align}
    \min_{\bm{w}^*, \bm{\mu}^*}\quad &C(\bm{w}^*):=\sum_{i=1}^nC_i(w_i^*)\,,\\
    \mathrm{s.t.} \quad &\ \bm{k_x}^{-1} (\bm{h}^{-1}(\Bar{y}\mathbbold{1}_n)) = \bm{w}^* -\bm{E}\bm{\mu}^*\label{subeq:opt_equlity2}, 
\end{align}
\end{subequations}
where $\bm{w}^*$ is the unique minimizer if and only if it satisfies $\nabla C_i (w_i^*) = \nabla C_j (w_j^*)\,,\forall i, j\in\mathcal{V}.$
 \end{lemma}


 \begin{proof}
We prove the equivalence of optimization by showing that the  steady-state output $\bm{y}^*=\Bar{y}\mathbbold{1}_n$ if and only if $\bm{k_x}^{-1} (\bm{h}^{-1}(\Bar{y}\mathbbold{1}_n)) = \bm{w}^*-\bm{E}\bm{\mu}^*$.
 First, we show necessity. If $\bm{y}^*=\Bar{y}\mathbbold{1}_n$, then the bijection in Assumption~\ref{ass:mapping} gives the unique input  $\bm{u}^*=\bm{k_x}^{-1} (\bm{h}^{-1}(\Bar{y}\mathbbold{1}_n)) $. Thus, $\bm{k_x}^{-1} (\bm{h}^{-1}(\Bar{y}\mathbbold{1}_n))=\bm{u}^*=\bm{w}^*-\bm{E}\bm{\mu^*} $.  
 Next, we show sufficiency. If $\bm{k_x}^{-1} (\bm{h}^{-1}(\Bar{y}\mathbbold{1}_n)) = \bm{w}^*-\bm{E}\bm{\mu}^*$, we have $\bm{u}^*=\bm{w}^*-\bm{E}\bm{\mu}^*=\bm{k_x}^{-1} (\bm{h}^{-1}(\Bar{y}\mathbbold{1}_n)) $. Then the bijection in Assumption~\ref{ass:mapping} gives $\bm{y}^*=\Bar{y}\mathbbold{1}_n$.

To prove that the marginal costs are identical at optimality, consider the Lagrangian function 
$
\mathcal{L}(\bm{w}^*,\bm{\mu}^*,\bm{\lambda}):=\sum_{i=1}^nC_i(w_i^*)+\bm{\lambda}^\top ( \bm{k_x}^{-1} (\bm{h}^{-1}(\Bar{y}\mathbbold{1}_n))-\bm{w}^* +\bm{E}\bm{\mu}^*)
$, where $\bm{\lambda}\in\real^n$ is the multiplier.
Then, the Karush-Kuhn-Tucker conditions~\cite{Boyd2004convex} gives
\begin{subequations}
    \begin{align}
    &\!\!\!\!\nabla_{\bm{w}^*}\mathcal{L}(\bm{w}^*,\bm{\mu}^*,\bm{\lambda})\!=\!\nabla_{\bm{w}^*}C(\bm{w}^*)-\bm{\lambda}=\mathbbold{0}_n\,,\label{subeq:grad_C_w} \\
    &\!\!\!\!\nabla_{\bm{\mu}^*}\mathcal{L}(\bm{w}^*,\bm{\mu}^*,\bm{\lambda})=\bm{E}^\top\bm{\lambda}\!=\!\mathbbold{0}_m\label{subeq:grad_C_mu}\,, \\
    &\!\!\!\!\nabla_{\bm{\lambda}}\mathcal{L}(\bm{w}^*,\bm{\mu}^*,\bm{\lambda})\!=\!\bm{k_x}^{-1} (\bm{h}^{-1}(\Bar{y}\mathbbold{1}_n)) \!- \!\bm{w}^* \!+\!\bm{E}\bm{\mu}^*\!=\!\mathbbold{0}_n\,.\label{subeq:grad_C_lambda} 
    \end{align}
\end{subequations} By~\cite{biggs1993algebraic}, the incidence matrix of a connected graph satisfy $\mathcal{N}(\bm{E}^\top) =\text{range}\left\{\mathbbold{1}_n\right\} $.
Hence,~\eqref{subeq:grad_C_mu} implies $\bm{\lambda}\in \text{range}\left\{\mathbbold{1}_n\right\}$ and~\eqref{subeq:grad_C_w} further yields $\nabla_{\bm{w}^*}C(\bm{w}^*)=\bm{\lambda}\in \text{range}\left\{\mathbbold{1}_n\right\}$. Moreover,~\eqref{subeq:grad_C_lambda} is satisfied since $\bm{w}^*$ and $\bm{\mu}^*$ are variables at the equilibrium. Hence, $\bm{w}^*$ solves~\eqref{eq: optimization_equivalent} if and only if $\nabla_{\bm{w}^*}C(\bm{w}^*)\in \text{range}\left\{\mathbbold{1}_n\right\}$, which concludes the proof.
 \end{proof}


By Lemma~\ref{lem: Equivalent_opf}, enforcing that $\bm{w}^*$ achieves identical marginal cost, i.e., $\nabla C_i (w_i^*) = \nabla C_j (w_j^*),\forall i, j\in\mathcal{V}$ can ensure that the steady-state actions settle down to the
solution of the resource allocation problem. To this end,
prior works have designed distributed
averaging-based integral control by communicating $\nabla C_i(w_i)$ with its neighbours~\cite{Zhao2015acc,andreasson2014distributed, weitenberg2018exponential}. However, they are restricted to quadratic costs and linear controllers. In this paper, we consider nonlinear controllers and a more general class of cost functions in the following assumption.

\begin{assumption}[Scaled-cost gradient functions]\label{ass: cost-grad-scale} The function $C_i(\cdot):\real\mapsto \real$ is strictly convex and continuously differentiable for all $i\in\mathcal{V}$.  Moreover, there exists a function $C_\mathrm{o}(\cdot):\real\mapsto \real$ and a group of positive scaling factors $\bm{c}:=\left(c_i, i \in \mathcal{V} \right)$ such that $\nabla C_i(\cdot)=\nabla C_\mathrm{o}(c_i \cdot), \forall i\in\mathcal{V}\,$.
\end{assumption}

Some examples satisfying Assumption~\ref{ass: cost-grad-scale} are 
1) \emph{polynomials of the form: }
$C_i(w_i)= \frac{c_i}{p} w_i^p+b_i$
where $c_i>0$ and $p$ is an even integer (this includes quadratics). 2) \emph{functions that are identical up to constants:} $C_i(w_i)= C_\mathrm{o}(w_i)+b_i$ (e.g., power generators of the same type but with different startup costs). 


%

\subsection{Structured controller design}
We aim to design the control law such that the control effort reaches the solution of  the optimal resource allocation problem~\eqref{eq: allocation_optimization} at the steady state, which can be equivalently realized through identical marginal cost at the steady state by Lemma~\ref{lem: Equivalent_opf}. Hence, we design the mechanism such that neighbouring nodes communicate their marginal cost and reach the consensus at the steady state.
We model communication network within the physical networked system 
as a connected graph $\tilde{\mathcal{G}} =(\mathcal{V},\tilde{\mathcal{E}} )$ with an incidence matrix $\tilde{\bm{E}}$. By adding the communication loop into the integral variable $\bm{s}$, the integral control term $\bm{r}(\bm{s})$ can respond to the difference of marginal costs. The edges in $\tilde{\mathcal{E}}$ are not necessarily the same as $\mathcal{E}$ and we use $\tilde{}$ to denote all variables belonging to the edges in the communication graph. 
The communication network associated with nodes of the physical network is designed as follows 
\begin{subequations}\label{eq: communication_network}
\begin{align}
\begin{split}
\mathcal{V}_{i}: \quad
\dot {s}_{i} &=-(y_i-\Bar{y})-c_iq_i, \\
\quad q_i &= 
     \sum_{l=1}^m \tilde{\bm{E}}_{i,l}\tilde{\mu}_{l},\\
     \quad o_i &=\nabla C_i(r_i(s_i))
   , \quad i \in \mathcal{V} \\   
\end{split}
    \\
   \begin{split}
    \quad &\tilde{\mathcal{E}}_{l}: \quad \tilde{\zeta}_{l} = o_i-o_j,\\
    &\quad\quad\quad \tilde{\mu}_{l}=\phi_{l}\left(\tilde{\zeta}_{l}\right), \quad l=(i,j)\in  \tilde{\mathcal{E}}       
    \end{split}
\end{align}    
\end{subequations}
with state $s_i\in \real$, input $q_i\in\real$, output $o_i\in\real$ for the nodes, and input $\tilde{\zeta}_{l}\in\real$, output $\tilde{\mu}_{l}\in\real$ for the edges. Note that the edges are designed as a memoryless system without states. Compactly, we have the closed-loop dynamics for the communication graph represented by $\dot{\bm{s}}=-(\bm{y}-\Bar{y}\mathbbold{1}_n)-\hat{\bm{c}} \tilde{\bm{E}} \bm{\phi}\left(\tilde{\bm{E}}^\top\nabla \bm{C}(\bm{r}(\bm{s}))\right)$, where $\hat{\bm{c}}=\operatorname{diag}(c_1,\cdots,c_n).$ Then, the control law is designed as follows.


\begin{design}[Distributed Steady-State Optimization]\label{design: optimal_flow}
For each node $i\in\mathcal{V}$, the external control law is $w_i=p_i(\Bar{y}_i-y_i)+r_i(s_i)$, where $p_i(\cdot):\mathbb{R}\mapsto\mathbb{R}$ and  $r_i(\cdot):\mathbb{R}\mapsto\mathbb{R}$ are Lipschitz continuous and strictly increasing functions with $p_i(0)=0$, $r_i(0)=0$. The functions $\psi_{l}(\cdot)$ are strictly increasing for all $l\in\mathcal{E}$. The ancillary state $\bm{s}$ comes from the communication network~\eqref{eq: communication_network} where the function $\phi_{l} (z):\real\mapsto \real $ is an  odd function and with the same sign as $z$ for all $l\in\tilde{\mathcal{E}}$.   Compactly,  we have
\begin{subequations}\label{eq: PIcontroller_cost}
\begin{align}
    \bm{u} &=-\bm{E}\bm{\psi}(\bm{\eta}) + \bm{p}(-\bm{y}+\Bar{y}\mathbbold{1}_n) +  \bm{r}(\bm{s})\\
    \dot{\bm{s}} &=-(\bm{y}-\Bar{y}\mathbbold{1}_n)\underbrace{-\hat{\bm{c}} \tilde{\bm{E}} \bm{\phi}\left(\tilde{\bm{E}}^\top\nabla \bm{C}(\bm{r}(\bm{s}))\right)}_{\text{the term added to } \eqref{eq: PIcontroller1}.}\label{subeq:s_opf}
\end{align}    
\end{subequations}
\end{design}



The following lemma shows properties of the added term in~\eqref{eq: PIcontroller_cost}, providing an intuition about why Controller Design~\ref{design: optimal_flow} can guarantee identical marginal cost (i.e., $\nabla C(\bm{r}(\bm{s}^*))\in\range{\mathbbold{1}_n}$).


\begin{lemma}[Cross term in the communication network]\label{lem:sign-cross}
Suppose Assumption~\ref{ass: cost-grad-scale} holds. 
Then 
\begin{equation}\label{eq:cross_communication}
\hat{\bm{c}}\tilde{\bm{E}}\bm{\phi}\left(\tilde{\bm{E}}^\top\nabla \bm{C}\left(\bm{r}\left(\bm{s}\right)\right)\right)=\mathbbold{0}_n    
\end{equation}
if and only if $\nabla \bm{C}(\bm{r}(\bm{s}))\in\range{\mathbbold{1}_n}$. Moreover, $\bm{r}(\bm{s})^T \hat{\bm{c}}\tilde{\bm{E}}\bm{\phi}\left(\tilde{\bm{E}}^\top\nabla \bm{C}\left(\bm{r}\left(\bm{s}\right)\right)\right)\geq0$
with equality holds if and only if $\nabla \bm{C}(\bm{r}(\bm{s}))\in\range{\mathbbold{1}_n}$.
\end{lemma}
The proof is given in Appendix~\ref{app:lem_sign-cross} by expanding the terms and  the properties of cost functions satisfying Assumption~\ref{ass: cost-grad-scale}. In particular, we use the fact that $\phi_l(\cdot)$ is an odd function.
We will show in the next subsection that $\bm{y}^*=\Bar{y}\mathbbold{1}_n$ is maintained and thus  $\bm{w}^*=\bm{r}(\bm{s}^*)$. Then, $\nabla \bm{C}(\bm{r}(\bm{s}^*))\in\range{\mathbbold{1}_n}$ is equivalent to $\nabla \bm{C}(\bm{w}^*)\in\range{\mathbbold{1}_n}$.

\subsection{Unique equilibrium with steady-state optimality}
The next theorem states that the closed-loop system~\eqref{eq:node_dyn}-\eqref{eq:couple} with Controller Design~\ref{design: optimal_flow} yields a unique equilibrium that guarantees output agreement and optimal resources allocation at the steady state.

\begin{theorem}[Steady-state optimality]\label{thm:equilibrium_cost}
Suppose Assumptions~\ref{ass: EIP_node}-\ref{ass: cost-grad-scale} hold and the input $\bm{u}$ follows~\eqref{eq: PIcontroller_cost} in the Controller Design~\ref{design: optimal_flow},
then the equilibrium is uniquely characterized by
\begin{subequations}\label{eq:equ-DAI} 
\begin{align}
\bm{y}^*=&\ \Bar{y}\mathbbold{1}_n, \;\;\\
\bm{x}^*=& \bm{h}^{-1}\left(\Bar{y}\mathbbold{1}_n\right)\,,\\ 
\bm{r}\left(\bm{s}^*\right) =&\ \nabla C_\mathrm{o}^{-1}(\gamma)\hat{\bm{c}}^{-1}\mathbbold{1}_n\,, \label{eq:equ-DAI-u}\\ 
-\bm{E}\bm{\psi}(\bm{\eta}^*)=&\bm{k_x}^{-1} (\bm{h}^{-1}(\Bar{y}\mathbbold{1}_n))-\nabla C_\mathrm{o}^{-1}(\gamma)\hat{\bm{c}}^{-1}\mathbbold{1}_n \,,\label{eq:equ-DAI-delta}
\end{align}
\end{subequations}
where  $\nabla C_\mathrm{o}^{-1}(\cdot)$ is the inverse of $\nabla C_\mathrm{o}(\cdot)$ and $\gamma$ is the unique solution to 
\begin{equation}\label{eq:gamma}
\nabla C_\mathrm{o}^{-1}(\gamma)=-\left(\mathbbold{1}_n^\top\bm{k_x}^{-1} (\bm{h}^{-1}(\Bar{y}\mathbbold{1}_n)) \right)/\left(\sum_{i=1}^n c_i^{-1}\right). 
\end{equation}

In particular, $\bm{w}^*=\bm{r}(\bm{s}^*)$ and
$\nabla C_i (w_i^*) = \nabla C_j (w_j^*)\,,\forall i,j\in\mathcal{V}\,.$ That is,  $\bm{w}^*$  at
the equilibrium solves the optimal resource allocation problem~\eqref{eq: optimization_equivalent}.
\end{theorem}
The proof is given in Appendix~\ref{app:thm_equilibrium_cost}. The key steps follow the equality at equilibrium and conditions in Lemma~\ref{lem:sign-cross}. 

\subsection{Asymptotic stability guarantees}
The next theorem shows that the unique equilibrium achieved by the Controller Design~\ref{design: optimal_flow} is locally asymptotically stable. 
\begin{theorem}[Stability]\label{thm:equilibrium_cost_stab}
Suppose assumptions in Theorem~\ref{thm:equilibrium_cost} hold. The closed-loop system~\eqref{eq:node_dyn}-\eqref{eq:edge_dyn}  is locally asymptotically stable at the unique equilibrium characterized by~\eqref{eq:equ-DAI}.
\end{theorem}

We prove that the equilibrium is asymptotically stable by constructing a  Lyapunov function the same as~\eqref{eq:Lyapunov_function}. Compared with~\eqref{eq:dotLyap1}, the added term in~\eqref{eq: PIcontroller_cost} creates the term  $\left(\bm{r}(\bm{s})-\bm{r}\left(\bm{s}^*\right)\right)^\top\left(-\hat{\bm{c}} \tilde{\bm{E}} \bm{\phi}\left(\tilde{\bm{E}}^\top\nabla \bm{C}(\bm{r}(\bm{s}))\right)\right)$ in $\dot{V}$.
The next Lemma shows that $\bm{r}\left(\bm{s}^*\right)^\top \hat{\bm{c}} \tilde{\bm{E}} \bm{\phi}\left(\tilde{\bm{E}}^\top\nabla \bm{C}(\bm{r}(\bm{s}))\right)$ in the extra term has no impact on the sign of $\dot{V}$.

\begin{lemma}\label{lem:sign-cross-cor}
Suppose Assumption~\ref{ass: cost-grad-scale} holds. 
Then 
\begin{equation}\label{eq:cross_communication_equalibrium}
\bm{r}\left(\bm{s}^*\right)^\top \hat{\bm{c}} \tilde{\bm{E}} \bm{\phi}\left(\tilde{\bm{E}}^\top\nabla \bm{C}(\bm{r}(\bm{s}))\right)=\mathbbold{0}_n.    
\end{equation}
\end{lemma}
\begin{proof}
Plugging in $\bm{r}\left(\bm{s}^*\right) = \nabla C_\mathrm{o}^{-1}(\gamma)\hat{\bm{c}}^{-1}\mathbbold{1}_n$ in~\eqref{eq:equ-DAI-u} gives
\begin{align*}
&\bm{r}\left(\bm{s}^*\right)^\top \hat{\bm{c}} \tilde{\bm{E}} \bm{\phi}\left(\tilde{\bm{E}}^\top\nabla \bm{C}(\bm{r}(\bm{s}))\right) \\
&=\ \nabla C_\mathrm{o}^{-1}(\gamma)\mathbbold{1}_n^\top (\hat{\bm{c}}^{-1})^\top
\hat{\bm{c}}  \tilde{\bm{E}} \bm{\phi}\left(\tilde{\bm{E}}^\top\nabla \bm{C}(\bm{r}(\bm{s}))\right)\\
&= \ \nabla C_\mathrm{o}^{-1}(\gamma)\mathbbold{1}_n^\top   \tilde{\bm{E}} \bm{\phi}\left(\tilde{\bm{E}}^\top\nabla \bm{C}(\bm{r}(\bm{s}))\right),
\end{align*}
which equals to $\mathbbold{0}_n $ since $\mathbbold{1}_n^\top   \tilde{\bm{E}}=\mathbbold{0}_n$.
\end{proof}

 By Lemma~\ref{lem:sign-cross}, we have $-\bm{r}(\bm{s})^T \hat{\bm{c}}\tilde{\bm{E}}\bm{\phi}\left(\tilde{\bm{E}}^\top\nabla \bm{C}\left(\bm{r}\left(\bm{s}\right)\right)\right)\leq0$. Thus, the extra term does not affect the negative definiteness of $\dot{V}$ and therefore the controller with the communication network in~\eqref{eq: PIcontroller_cost} still maintains the stability of the system. The full proof of Theorem~\ref{thm:equilibrium_cost_stab} is given below.
\begin{proof}
We prove that the equilibrium is asymptotically stable by constructing a  Lyapunov function the same as~\eqref{eq:Lyapunov_function}:
\begin{equation}\label{eq:Lyapunov_function2}
\begin{split}
    &V_2(\bm{x},\bm{\eta},\bm{s} )|_{\bm{x}^*,\bm{\eta}^*,\bm{s}^*}\\ 
    &= \sum_{i=1}^{n} W_{i}^\mathcal{V}(x_i,x_i^*)  +B^\mathcal{E}(\bm{\eta},\bm{\eta}^*)+B^\mathcal{V}(\bm{s}, \bm{s}^*),    
\end{split}
\end{equation}
where the positive definiteness of $V_2(\bm{x},\bm{\eta},\bm{s} )|_{\bm{x}^*,\bm{\eta}^*,\bm{s}^*}$
 still holds because of the monotonicity of the functions $r_i(\cdot)$ and $\psi(\cdot)$ for all $i$ in Controller Design~\ref{design: optimal_flow}. Hence, $V_2(\bm{x},\bm{\eta},\bm{s} )|_{\bm{x}^*,\bm{\eta}^*,\bm{s}^*}$ is a well-defined Lyapunov function.

The time derivative of $B^\mathcal{V}(\bm{s}, \bm{s}^*)$ is
\begin{align}
    &\dot{B}^\mathcal{V}(\bm{s}, \bm{s}^*) \label{eq:dot_bregman_s_comm}\\ 
    &=\left(\nabla R \left(\bm{s}\right)-\nabla R \left(\bm{s}^*\right)\right)^\top \dot{\bm{s}} \nonumber \\
    & \stackrel{\circled{1}}{=}\left(\bm{r}(\bm{s})\!-\!\bm{r}\left(\bm{s}^*\right)\right)^\top\left(-(\bm{y}-\bm{y}^*)-\hat{\bm{c}} \tilde{\bm{E}} \bm{\phi}\left(\tilde{\bm{E}}^\top\nabla \bm{C}(\bm{r}(\bm{s}))\right)\right)\!,  \nonumber
\end{align}
where $\circled{1}$ follows from $\nabla R \left(\bm{s}\right)=\bm{r}\left(\bm{s}\right)$ and $\dot{\bm{s}}=\left(\!-(\bm{y}\!-\!\bm{y}^*)\!-\!\hat{\bm{c}} \tilde{\bm{E}} \bm{\phi}\left(\tilde{\bm{E}}^\top\nabla \bm{C}(\bm{r}(\bm{s}))\right)\right)$ by Controller Design~\ref{design: optimal_flow}.

The time derivative of the Lyapunov function in~\eqref{eq:Lyapunov_function2} is
\begin{equation}\label{eq:dotLyap2}
\begin{split}
    &\dot{V}_2(\bm{x},\bm{\eta},\bm{s})|_{\bm{x}^*,\bm{\eta}^*,\bm{s}^*}  
    \\
    &= \sum_{i=1}^{n} \dot{W}_{i}^\mathcal{V}\left(x_{i}, x_{i}^*\right) + \dot{B}^\mathcal{E}(\bm{\eta},\bm{\eta}^*)+\dot{B}^\mathcal{V}(\bm{s}, \bm{s}^*)
    \\
    &
    \begin{split}
     \stackrel{\circled{1}}{=} 
     &\dot{V}_1(\bm{x},\bm{\eta},\bm{s})|_{\bm{x}^*,\bm{\eta}^*,\bm{s}^*}  \\
      &+\left(\bm{r}(\bm{s})-\bm{r}\left(\bm{s}^*\right)\right)^\top\left(-\hat{\bm{c}} \tilde{\bm{E}} \bm{\phi}\left(\tilde{\bm{E}}^\top\nabla \bm{C}(\bm{r}(\bm{s}))\right)\right)  
    \end{split}
   \\
      &\stackrel{\circled{2}}{=}\dot{V}_1(\bm{x},\bm{\eta},\bm{s})|_{\bm{x}^*,\bm{\eta}^*,\bm{s}^*}\\
      &\quad-\bm{r}(\bm{s})^\top\left(\hat{\bm{c}} \tilde{\bm{E}} \bm{\phi}\left(\tilde{\bm{E}}^\top\nabla \bm{C}(\bm{r}(\bm{s}))\right)\right) 
         \\
      &\stackrel{\circled{3}}{\leq}\dot{V}_1(\bm{x},\bm{\eta},\bm{s})|_{\bm{x}^*,\bm{\eta}^*,\bm{s}^*}
   \stackrel{\circled{4}}{\leq} -\sum_{i=1}^{n} \rho_{i}\left\|y_{i}-y_{i}^*\right\|^{2}
\end{split}
\end{equation}
where the equality $\circled{1}$ uses~\eqref{eq:dot_bregman_s_comm}. The equality $\circled{2}$ uses $\bm{r}\left(\bm{s}^*\right)^\top \hat{\bm{c}} \tilde{\bm{E}} \bm{\phi}\left(\tilde{\bm{E}}^\top\nabla \bm{C}(\bm{r}(\bm{s}))\right)=\mathbbold{0}_n$ in Lemma~\ref{lem:sign-cross-cor}.
The inequality $\circled{3}$ uses $\bm{r}(\bm{s})^\top\left(\hat{\bm{c}} \tilde{\bm{E}} \bm{\phi}\left(\tilde{\bm{E}}^\top\nabla \bm{C}(\bm{r}(\bm{s}))\right)\right) \geq0$ in Lemma~\ref{lem:sign-cross}. The inequality $\circled{4}$ follows directly from~\eqref{eq:dotLyap1}.

Therefore, $\dot{V}_2(\bm{x},\bm{\eta},\bm{s})|_{\bm{x}^*,\bm{\eta}^*,\bm{s}^*} \leq0$ with equality only holds at the  equilibrium. By Lyapunov  conditions, the system is locally asymptotically stable around the equilibrium. 
\end{proof}

\section{Monotone Neural Network Controller }
\label{sec:RL}

\begin{figure*}[ht]	
	\centering	\includegraphics[width=0.9\textwidth]{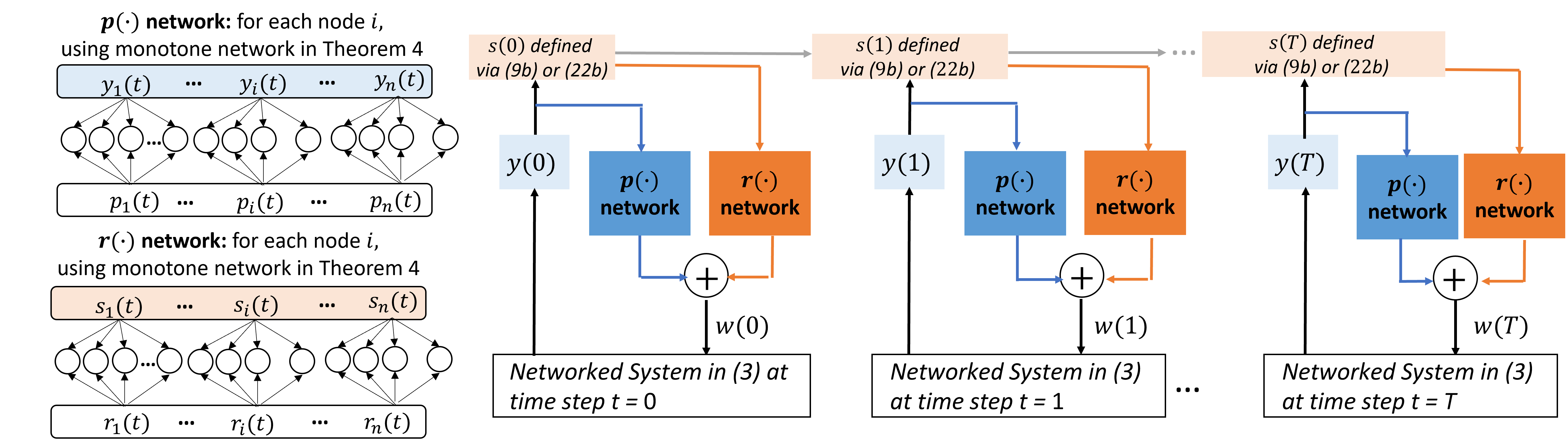}
	\caption{ Computation Graph for Training the Neural PI Controllers. \vspace{-0.3cm} 
	}	\label{fig:computation_graph_training}
\end{figure*}

 Controller Design~\ref{design: agreement_set} and~\ref{design: optimal_flow} demonstrate the stability and steady-state guarantees achieved by  the   structures with monotonicity conditions, but finding and optimizing the monotone functions remains a challenge. 
In this section, we show how to parameterize neural networks such that the monotonicity conditions can be met. On this basis, the functions in Controller Design~\ref{design: agreement_set} and~\ref{design: optimal_flow} can be parameterized through the monotone neural networks and optimized through training to improve the transient performances. 

\subsection{Monotone neural network design} \label{sec:increasing_NN}

Controller Design~\ref{design: agreement_set} and~\ref{design: optimal_flow} require functions that are  monotonically increasing and cross the origin. 
We parameterize this class of functions using the monotone neural network we proposed in~\cite{cui2022tps}. In this paper, we rigorously prove a universal approximation result: any increasing functions through the origin can be approximated by this construction.

\begin{theorem}[Universal approximation of monotonic functions \label{thm:stack_relu}]
Let $\bm{\alpha^{+(-)}}\in\real^d$ and $\bm{\beta^{+(-)} }\in\real^d$ be the weights and biases for a single-hidden-layer neural network with $d$ neurons. The activation is the ReLU  function defined by $\bm{\sigma}(\bm{z})=\max(\bm{z},0)$ for $\bm{z}\in\real^d$ with $\max(\cdot)$ denotes the element-wise maximization.
Let $\mathcal{Z}$ be a closed interval in $\real$ and $r(z):\mathcal{Z}\mapsto\real$ be a bounded, Lipschitz continuous and strictly increasing function through the origin.
For any $\epsilon>0$, there exists a function $g(z):\mathcal{Z}\mapsto\real$ constructed by
\begin{subequations}\label{eq:stacked-relu}
\begin{align}
     g(z)= &\left(\bm{\alpha^+ }\right)^\top\!\bm{\sigma}(\mathbbold{1}_d z\!-\!\bm{\beta }^+)\!+\!
     \left(\bm{\alpha^- }\right)^\top\!\bm{\sigma}(-\mathbbold{1}_d z\!+\!\bm{\beta }^-)\\
     \mbox{where }
     &-\infty<\sum_{j=1}^{d'} \alpha_j^-<0<\sum_{j=1}^{d'} \alpha_j^+<\infty\label{subeq:relu_weights} \,, \forall d' = 1, ..., d\\
     &\beta_d^-\leq\cdots\leq \beta_1^-=0=\beta_1^+\leq\cdots\leq \beta_d^+\label{subeq:relu_bias},
\end{align}
\end{subequations}
such that $\left|r(z)-g(z)\right|<\epsilon$ for all $z \in \mathcal{Z}$. 
\end{theorem}
To prove the universal approximation property in Theorem~\ref{thm:stack_relu}, we first show that piecewise linear functions with positive slopes can approximate all monotonically increasing functions, then explicitly construct a monotone neural network to represent the piecewise linear functions. The proof of Theorem~\ref{thm:stack_relu} can be found in Appendix~\ref{app:lemma_stack_relu }.
  \begin{remark}
  The function $g(z):\mathcal{Z}\mapsto\real$ constructed by~\eqref{eq:stacked-relu} is a piecewise linear function where the slope of each piece is positive. The smallest slope is $\gamma = \min\left\{\min_{d' = 1, ..., d} \sum_{j=1}^{d'} \alpha_j^+ , \min_{d' = 1, ..., d} |\sum_{j=1}^{d'} \alpha_j^-|\right\}$. Hence, for all $z_1,z_2 \in \mathcal{Z}$, we have
 $\left(g(z_1)-g(z_2)\right)\left(z_1-z_2\right)\geq \gamma\left(z_1-z_2\right)^2$  
 and thus $g(z)$ is also strongly increasing.
  \end{remark}
  

\subsection{Implementation of the monotone neural network in Theorem~\ref{thm:stack_relu}}\label{app: trick_relu}
The constraints in~\eqref{subeq:relu_weights} and~\eqref{subeq:relu_bias}  are not trivial to enforce when training the neural network. As 
 we briefly mentioned in our previous works~\cite{cui2022tps, shi2021stability}, the constraints in~\eqref{subeq:relu_weights} and~\eqref{subeq:relu_bias} can be automatically satisfied by a simple re-parameterization of parameters. Specifically, we introduce a group of nonzero intermediate parameters  $\tilde{\bm{\alpha}}^{+(-)}\in\real^d$ and $\tilde{\bm{\beta}}^{+(-)} \in\real^{d-1}$ that are unconstrained such that the original parameters  $\bm{\alpha^{+(-)}}\in\real^{d}$ and $\bm{\beta^{+(-)} }\in\real^d$ are parameterized as
\begin{equation}\label{eq: relu_trick}
    \begin{split}
        &\alpha_1^+=\left(\tilde{\alpha}_1^+\right)^2, \quad \alpha_j^+=\left(\tilde{\alpha}_j^+\right)^2\!-\!\left(\tilde{\alpha}_{j-1}^+\right)^2\!, \forall j=2,\cdots\!,d
        \\
        &\alpha_1^-\!=\!-\!\left(\tilde{\alpha}_1^-\right)^2\!, \quad \alpha_j^-\!=\!-\left(\tilde{\alpha}_j^-\right)^2\!+\!\left(\tilde{\alpha}_{j-1}^-\right)^2, \forall j=2,\cdots\!,d \\
        &\beta_1^+=0, \quad \beta_j^+=\sum_{i=1}^{j-1}\left(\tilde{\beta}_i^+\right)^2,
        \forall j=2,\cdots,d
        \\
        &\beta_1^-=0, \quad\beta_j^-=-\sum_{i=1}^{j-1}\left(\tilde{\beta}_i^-\right)^2,
        \forall j=2,\cdots,d.
    \end{split}
\end{equation}
By inspection,~\eqref{subeq:relu_weights} and~\eqref{subeq:relu_bias}  naturally hold through the construction in~\eqref{eq: relu_trick}. 

Then the monotonic function in Section~\ref{sec:increasing_NN} is implemented using the pseudo-code  in Algorithm~\ref{alg:monotone}. It realizes a function $g(z)$ that is strictly increasing and crosses the origin. The trainable parameters are $\tilde{\bm{\alpha}}^{+}\in\real^d$, $\tilde{\bm{\alpha}}^{-}\in\real^d$, $\tilde{\bm{\beta}}^{+} \in\real^{d-1}$, $\tilde{\bm{\beta}}^{-} \in\real^{d-1}$. 
 \begin{algorithm}
 \caption{Monotone neural network implementation}
 \label{alg:monotone}
 \begin{algorithmic}[1]
 \renewcommand{\algorithmicrequire}{\textbf{Input:}}
 \renewcommand{\algorithmicensure}{\textbf{Output:}}
 \REQUIRE Variable $z$, 
nonzero weights $\tilde{\bm{\alpha}}^{+}\in\real^d$, $\tilde{\bm{\alpha}}^{-}\in\real^d$, $\tilde{\bm{\beta}}^{+} \in\real^{d-1}$, $\tilde{\bm{\beta}}^{-} \in\real^{d-1}$\\
Set up the value for the first neuron $\alpha_1^+=\left(\tilde{\alpha}_1^+\right)^2, \alpha_1^-=-\left(\tilde{\alpha}_1^-\right)^2,\beta_1^+=0, \beta_1^-=0$
  \FOR {$j = 2$ to $d$}
  \STATE  $\alpha_j^+=\left(\tilde{\alpha}_j^+\right)^2-\left(\tilde{\alpha}_{j-1}^+\right)^2$ , $\alpha_j^-=-\left(\tilde{\alpha}_j^-\right)^2+\left(\tilde{\alpha}_{j-1}^-\right)^2$\\
  \STATE 
  $\beta_j^+=\sum_{i=1}^{j-1}\left(\tilde{\beta}_i^+\right)^2$, $\quad\beta_i^-=-\sum_{i=1}^{j-1}\left(\tilde{\beta}_i^-\right)^2$
  \ENDFOR
    \ENSURE $g(z)= \left(\bm{\alpha^+ }\right)^\top\!\bm{\sigma}(\mathbbold{1}_d z\!-\!\bm{\beta }^+)
     \!+\!
     \left(\bm{\alpha^- }\right)^\top\!\bm{\sigma}(-\mathbbold{1}_d z+\bm{\beta }^-)$
 \end{algorithmic} 
 \end{algorithm}

\subsection{Optimizing transient performances through training} 
\label{sec:rnn_training}

Each monotonic function in the Controller Design~\ref{design: agreement_set} and Controller Design~\ref{design: optimal_flow} is parameterized by the neural network construction in Algorithm~\ref{alg:monotone}. Thus, stability and steady-state performances in Theorem~\ref{thm: output_level_stable} and~\ref{thm:equilibrium_cost_stab} are guaranteed through the construction by design. The transient performances are further optimized by training the neural networks, and most model-based or model-free learning algorithms can be utilized.

In Fig \ref{fig:computation_graph_training}, we use the training of the structured proportional controller $\bm{p}(\cdot)$ and the integral controller $\bm{r}(\cdot)$ as an example to visualize the detailed construction and the computation graph in the networked system defined in~\eqref{eq:couple}.
The  trainable parameters are contained in each node's $p_i(\cdot)$ and $r_i(\cdot)$ functions, where both are parameterized as monotone neural networks given in Algorithm~\ref{alg:monotone}. The signal $w_i(t)=p_i(t)+r_i(t)$ then serves  the external control  in the networked system defined in~\eqref{eq:couple} that evolves through time. 
 Let
$\bm{\Theta}_i$ be the  trainable parameters in neural networks of node $i$. The loss function is defined as $Loss(\bm{\Theta}) = \sum_{t=0}^{T} \sum_{i=1}^{n} J_i (y_i(t) - \bar{y}, w_i(t))$, where $\bm{\Theta}:=\left(\bm{\Theta}_i, i\in[n]\right)$ is involved in the parameterization of $\bm{w}$ and $J_i$ is the transient cost function the same as~\eqref{eq:transient_cost}. The parameters $\bm{\Theta}$ are then trained by reinforcement learning algorithms using the defined  loss function.


\section{Experiments}
\label{sec:experiments}

\begin{figure*}[ht]
\centering
\subfloat[Edge Feedback]{\includegraphics[width=1.5
in]{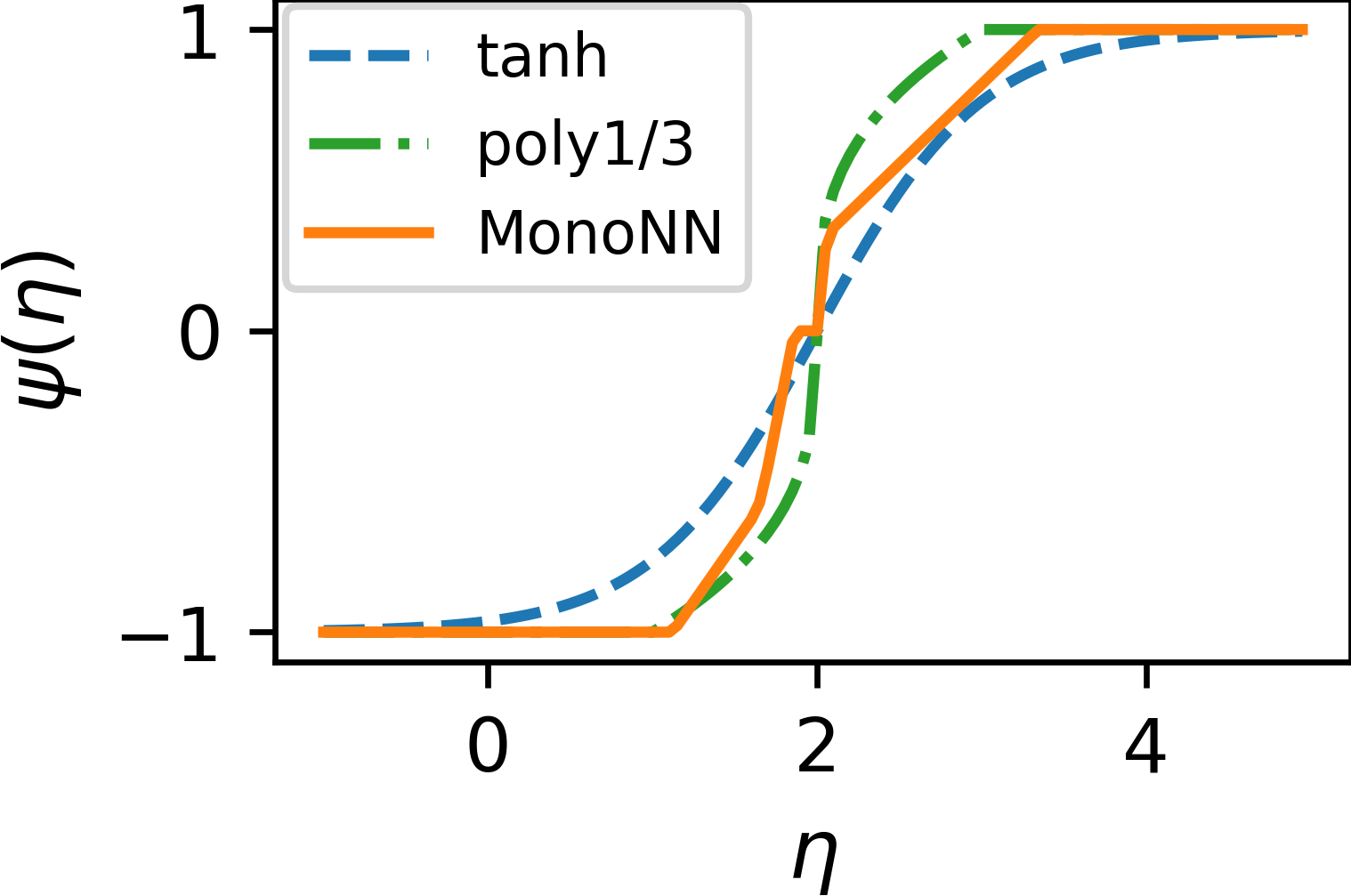}%
\label{fig_first_case}}
\subfloat[MonoNN]{\includegraphics[width=1.5in]{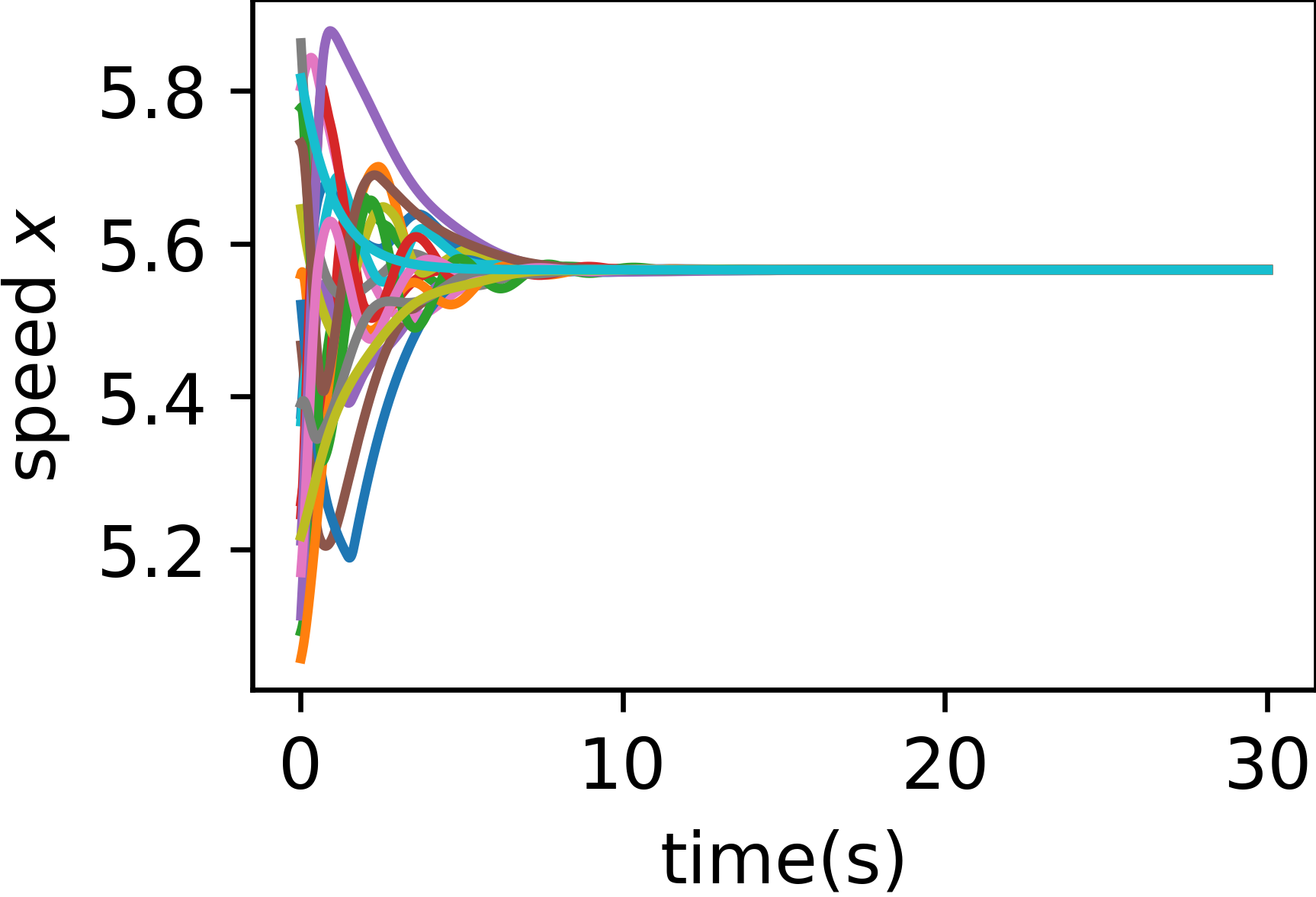}%
\label{fig_second_case_5}}
\subfloat[tanh]{\includegraphics[width=1.5in]{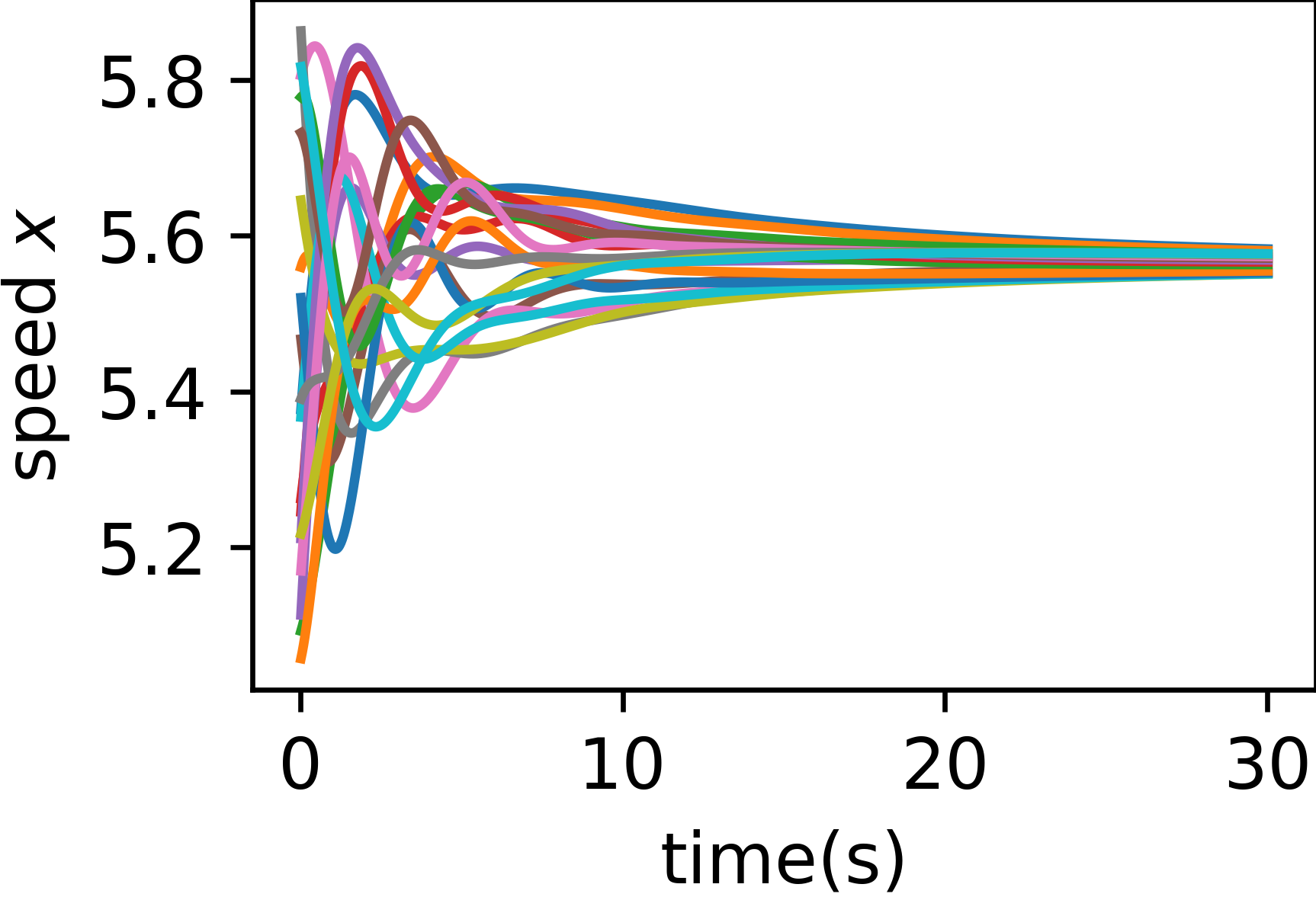}%
\label{fig_second_case_6}}
\subfloat[Poly1/3]{\includegraphics[width=1.5in]{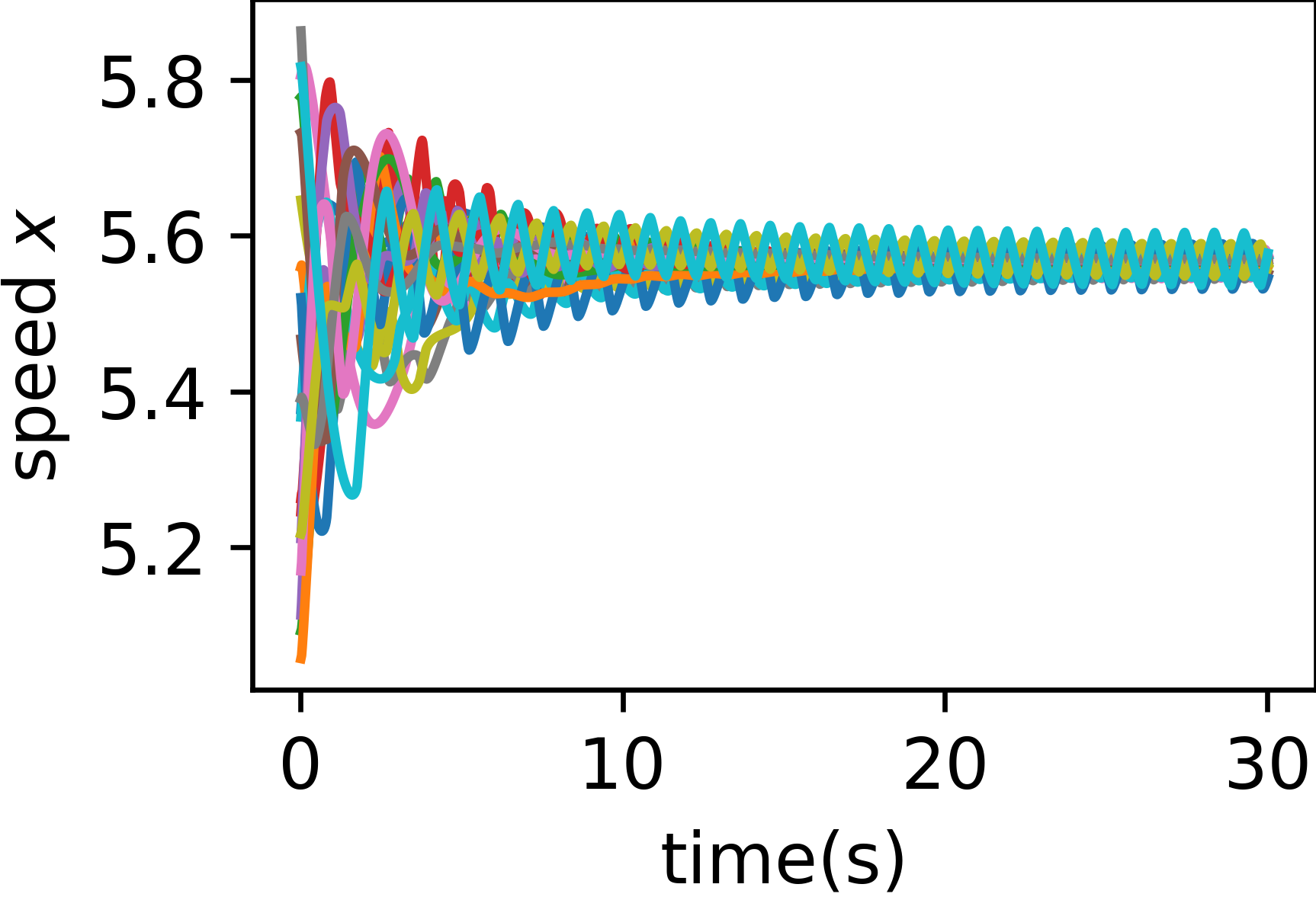}%
\label{fig_second_case_7}}
\caption{ (a) Comparison of the edge feedback functions learned by MonoNN with tanh and poly1/3.   (b)-(c) Dynamics of the system with the three edge feedback functions. All of the dynamics reach an agreement, with MonoNN converging quickest.\vspace{-0.4cm}}
\label{fig:Learn_edge_main}
\end{figure*}

We end the paper with case studies demonstrating the effectiveness of the proposed neural-PI controllers in two networked systems: vehicle platooning and power system frequency control. All experiments are run with a NVIDIA Tesla P100 GPU with 16GB memory. The proposed neural-PI controllers can be trained via most reinforcement learning algorithms, 
and we use the RNN-based algorithm in~\cite{cui2022tps,drgona2020learning} for training the neural network-based controllers.
Code for all experiments is available at
\url{https://github.com/Wenqi-Cui/NeuralPI_Networked_Systems}.

\subsection{Vehicle platooning}
\label{subsec: exp_traffic}
The first experiment is the vehicle platoon control in Fig.~\ref{fig:network}(b).
We adopt the model in~\cite{coogan2014dissipativity,burger2014duality}: (i) the drivers are heterogeneous and have different ``preferred" velocities, (ii) neighbouring vehicles influence each other through their relative distance,
and (iii) the set of neighbors to a vehicle remains unchanged within the simulation period. 
 
Let $\mathcal{V}$ be the set of all the vehicles.
The vehicle $i\in \mathcal{V}$ adjusts its velocity $x_{i}$ according to its preferred velocities $\lambda_{i}^{0}>0$ and the input $u_i$. 
The node dynamics are
\vspace{-0.2cm}
\begin{equation}\label{eq:traffic_dyn_node}
 \mathcal{V}_{i}: \quad \dot{x}_{i}=\kappa_{i}\left(-(x_{i}-\lambda_{i}^{0})+\frac{1}{\rho_i} u_{i}\right), \quad y_{i}=x_{i},   
\end{equation}
where $\kappa_{i}>0$ and $\rho_i>0$ are constants. The output $y_i$ is the velocity $x_{i}$ that can be observed by neighbouring vehicles. 


The neighbouring vehicles are described by edges in $\mathcal{E}$. If vehicle $i$ is a neighbor in front of vehicle $j$, we associate an edge $l$ with $i$ as the head and $j$ as the tail, written as $l=(i,j)\in \mathcal{E}$. 
For each edge $l$,   $\eta_l $ denotes the relative position of vehicles. The action of each node $i$ is $u_i=-\psi_{l} \left(\eta_l\right)+w_i$, where $\psi_{l} \left(\eta_l\right)$ is the feedback from the changes in the relative distance between neighbouring vehicles, and $w_i$ is the external input that synchronizes the outputs to $\Bar{y}$.

The edge dynamics are:
\vspace{-0.2cm}
\begin{equation}\label{eq:traffic_dyn_edge}
\mathcal{E}_{l}: \quad \dot{\eta}_{l}=\zeta_{l}, \quad \mu_{l}=\psi_{l}\left(\eta_{l}\right),
\end{equation}
where the input $\zeta_{l}$ is the relative velocity $x_{i}-x_{j}$ and the output $\mu_l$ is the position feedback $\psi_{l} \left(\eta_l\right)$. This recovers the input-output coupling of nodes and edges shown in Fig.~\ref{fig:network}. Appendix~\ref{app:assump_vehicle} shows that \eqref{eq:traffic_dyn_edge} satisfies Assumptions~\ref{ass: EIP_node} and \ref{ass:mapping}.

\subsubsection{Design freedom}
Vehicles in a platoon need to have the same velocity at the steady state to avoid collisions. The first design freedom is the edge feedback function $\psi_{l}(\cdot)$ and we follow the procedure in Controller Design~\ref{design: agreement_set}. The second  design freedom is the adjustments $w_i$ to the preferred velocity of vehicle $i$.
We would like to maintain at the required velocity $\Bar{y}$  at the lowest cost $\sum_{i\in \mathcal{V}}C_i(w_i)$. 


\subsubsection{Simulation setup}
We adopt the setup in~\cite{coogan2014dissipativity,burger2014duality}. The number of vehicles is $n=20$ and they are placed on a line. The sensitivity parameter is $\kappa_i=1$ for all vehicles. The parameters $\lambda_{i}^{0}$ and $\rho_i$ are randomly generated  by $\lambda_{i}^{0}\sim \texttt{uniform}[5,6]$ and $\rho_i\sim \texttt{uniform}[1,2]$, respectively. We generate 300 samples for training and testing, with initial velocities $x_i(0) \sim \texttt{uniform}[5,6]$. The state $\eta_l$ is initialized as 2 and $s_i$ is initialized as 0, respectively. The  stepsize in time is set as $\Delta t= 0.02s$ and for $K=300$ steps in a trajectory.  The communication graph is a randomly generated  regular graph with degree three.  
The episode number and batch size are 400 and 300,  respectively.


\subsubsection{Learning edge feedback functions}
We first demonstrate the performance of the learned edge feedback function $\bm{\psi}(\cdot)$ without the external control $\bm{w}$. References~\cite{coogan2014dissipativity} and~\cite{burger2014duality} 
 provide the algebraic constraint that $\bm{\psi}(\cdot)$ should be monotonically increasing. We parameterize $\bm{\psi}(\cdot)$ using the monotone neural network in~\eqref{eq:stacked-relu}. The loss function in training is set to be $J(\bm{y})=\sum_{i=1}^{n}\left(\sum_{k=200}^{K}|y_i(k\Delta t)-\frac{1}{n}\sum_{j=1}^{n}y_j(k\Delta t)|\right)+\sum_{l=1}^m\text{relu}(-\eta_l+1)$, which penalizes on the speed disagreement in the last 100 steps and the relative distance smaller than 1. This loss function  is used to encourage quicker convergence and avoid potential collisions. 
 We compare against the edge feedback functions presented in~\cite{coogan2014dissipativity,burger2014duality}, where $\psi_l(\eta_l) = \tanh(\eta_l)$ and $\psi_l(\eta_l) = (\eta_l)^{1/3}$ are used in~\cite{burger2014duality} and~\cite{coogan2014dissipativity}, respectively. These functions guarantee stabilization but without performance optimization.

Fig.~\ref{fig:Learn_edge_main} compares the performance of $\psi_l(\cdot)$ parameterized by the monotone neural network (labeled as MonoNN) in Algorithm~\ref{alg:monotone} with $\tanh(\eta_l-\eta_0)$ in~\cite{burger2014duality} and $(\eta_l-\eta_0)^{1/3}$(labeled as poly1/3) in~\cite{coogan2014dissipativity}, where $\eta_0=2$ is the initial distance of neighbouring vehicles. 
The shape of different edge feedback functions is shown in Fig.~\ref{fig:Learn_edge_main}(a).
Fig.~\ref{fig:Learn_edge_main}(b)-(d) visualize the transient velocity of all vehicles from the same initial condition. All of the dynamics reach an output agreement at approximate 5.6 m/s without external control $\bm{w}$, with MonoNN realizing better transient performance with much faster convergence. 


\subsubsection{Controller performance}
We implement the external control $\bm{w}$ to realize a specific output agreement at $\Bar{y}=5.2$ m/s and reduce the steady-state resource allocation cost. The transient cost is set to be $ J(\bm{y},\bm{w}) = \sum_{i=1}^n\sum_{k=1}^{K}| y_i(k\Delta t)-\Bar{y}|+c_i (w_i(k\Delta t))^2$, 
where $c_i\sim \texttt{uniform}[0.025,0.075]$. The steady-state cost in resource allocation~\eqref{eq: allocation_optimization} is $C(\bm{w}) = \sum_{i=1}^n c_i (w_i^*)^2$, where we use $w_i(20)$ to approximate $w_i^*$ since the dynamics approximately enter the steady state after $t=15s$ as we will show later in simulation.
The loss function in training is  $J(\bm{y},\bm{w})$, such that neural networks are optimized to reduce transient cost through training.

We compare the performance of the learned structured neural-PI controllers, 1) NeuralPI-Comm, the neural-PI controller with communication (Controller design~\ref{design: optimal_flow}) and, 2) NeuralPI-WoComm, the neural-PI controller without communication (Controller Design~\ref{design: agreement_set}). Both neural-PI controllers are parameterized by monotone neural networks, with 20 (i.e., $d=20$) neurons in the hidden layer. 
We compare against two benchmarks with communication: 
3) DenseNN-Comm: Two-layer dense neural networks with ReLU activation, with 20 neurons per hidden layer.  
4) LinearPI-Comm: Conventional linear PI control  parameterized by $p_i(\Bar{y}_i-y_i) = \theta_{i,1}(\Bar{y}_i-y_i)$ and $r(s_i) =\theta_{i,2}s_i $, where  $\theta_{i,1}$ and $\theta_{i,2}$ are linear coefficients optimized through learning.


\begin{figure}[ht]
\centering
\subfloat[Training Loss]{\includegraphics[width=1.7
in]{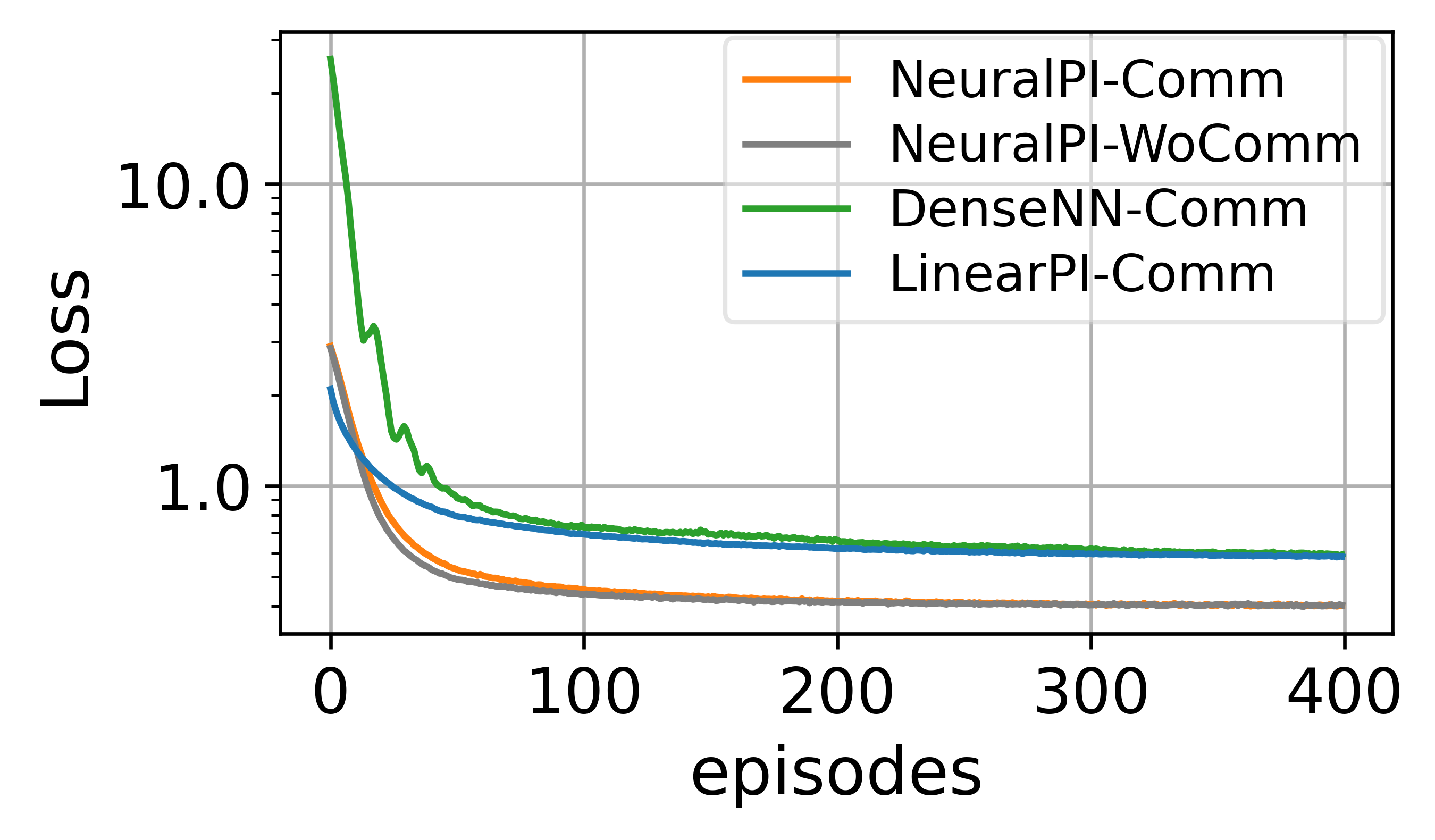}%
\label{subfig:Loss_epi_vehicel}}
\hfil
\subfloat[Average transient and steady cost ]{\includegraphics[width=1.7in]{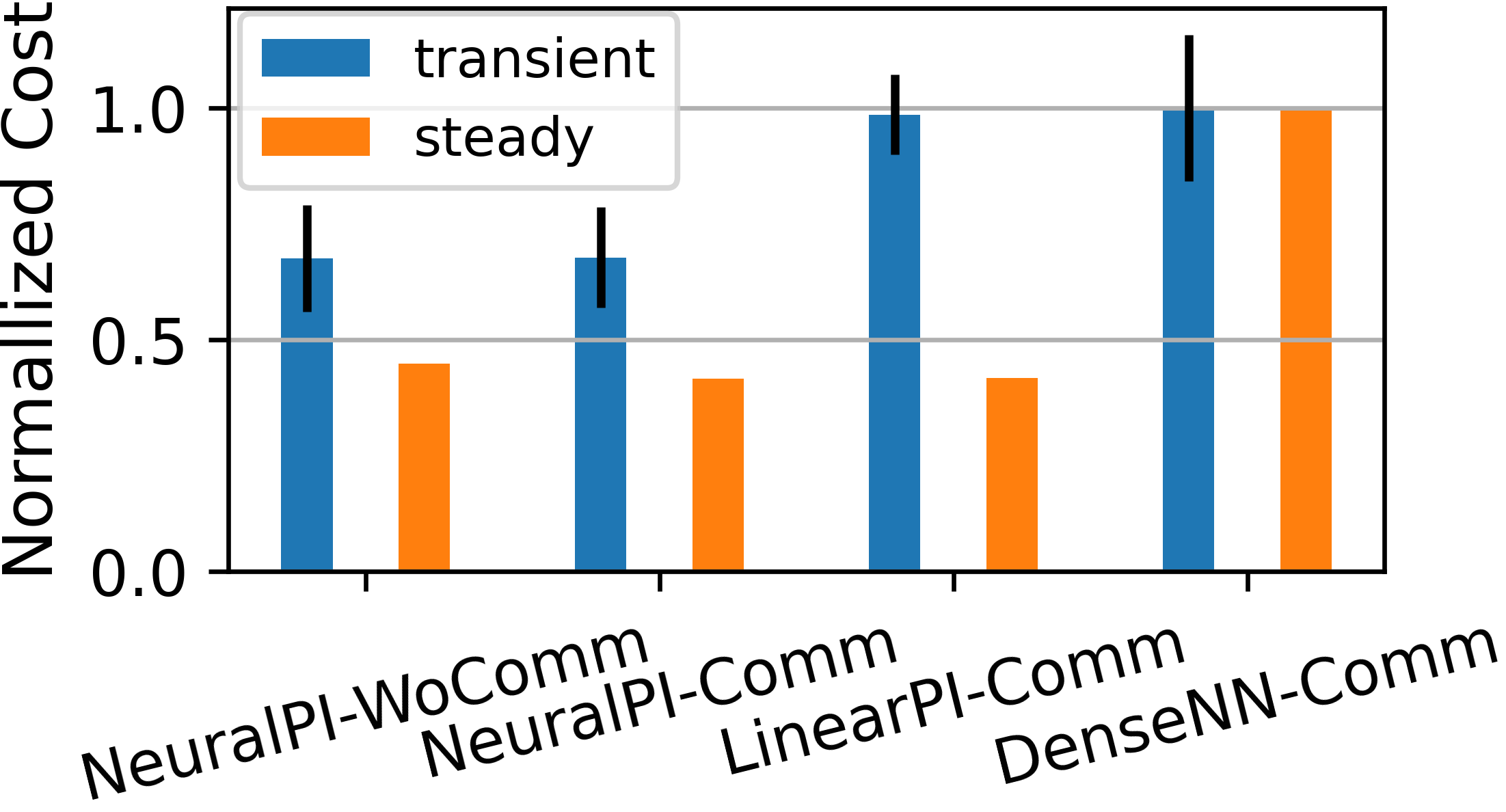}%
\label{fig_second_case_0}}
\caption{(a) Average batch loss along episodes. All converge, with the NeuralPI achieving the lowest cost. (b) The average transient cost and steady-state cost with error bar
on the randomly generated  test set with size 300. NeuralPI achieves  a transient cost that is much lower than others. 
NeuralPI-Comm and Linear-Comm lead to the same lowest steady-state cost guaranteed by Controller Design~\ref{design: optimal_flow}.\vspace{-0.4cm}}
\label{fig:Loss_epi_vehicel}
\end{figure}

The average batch loss  during episodes of training is shown in Fig.~\ref{fig:Loss_epi_vehicel}(a). All of the four methods converge, with the NeuralPI achieves the lowest cost.
Fig.~\ref{fig:Loss_epi_vehicel}(b) shows the transient and steady-state cost on the test set.   NeuralPI-Comm and LinearPI-Comm have the lowest possible steady-state cost, as guaranteed by Theorem~\ref{thm:equilibrium_cost}. NeuralPI also achieves a transient cost that is much lower than others. 

\begin{figure}[ht]
\centering
\subfloat[NeuralPI-Comm:  dynamics of $\bm{x}$ , $\nabla\bm{C}(\bm{r}(\bm{s}))$  and $w$ \vspace{-0.2cm}
]{\includegraphics[width=3.4
in]{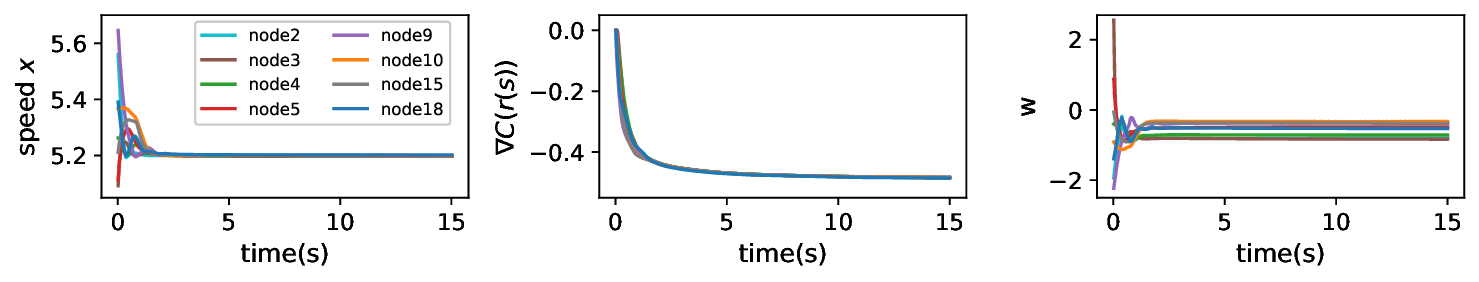}%
}
\hfil
\subfloat[NeuralPI-WoComm:  dynamics of $\bm{x}$ , $\nabla\bm{C}(\bm{r}(\bm{s}))$  and $w$ \vspace{-0.2cm}
]{\includegraphics[width=3.4
in]{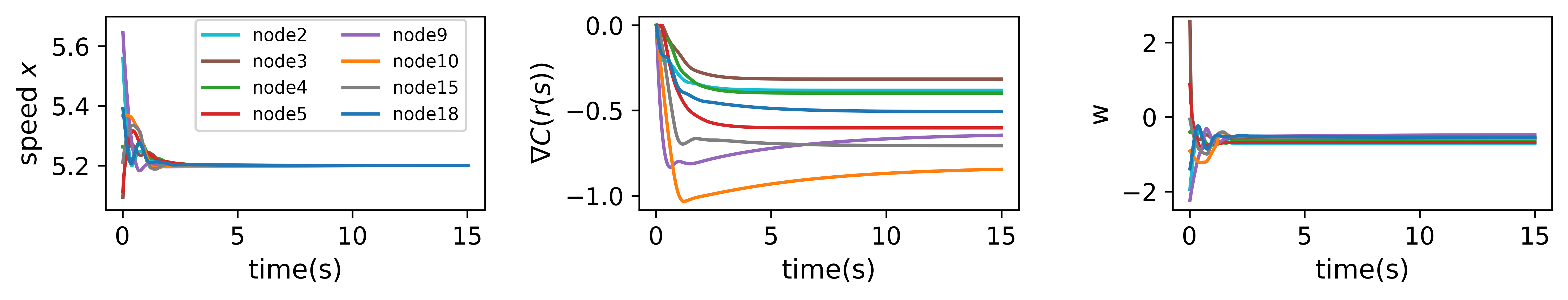}%
}
\hfil
\subfloat[LinearPI-Comm:
 dynamics of $\bm{x}$ , $\nabla\bm{C}(\bm{r}(\bm{s}))$  and $w$ \vspace{-0.2cm}
 ]{\includegraphics[width=3.4in]{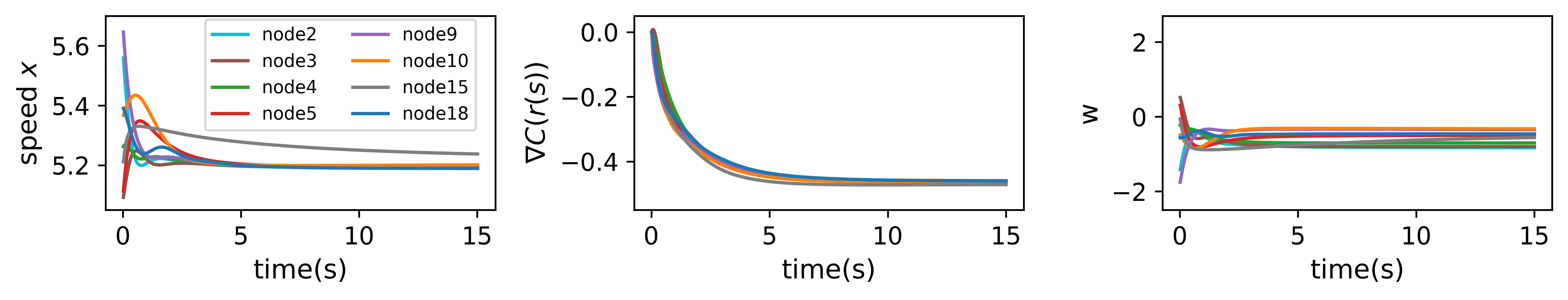}%
}
\hfil
\subfloat[DenseNN-Comm:  dynamics of $\bm{x}$ , $\nabla\bm{C}(\bm{r}(\bm{s}))$  and $w$ \vspace{-0.1cm}
]{\includegraphics[width=3.4in]{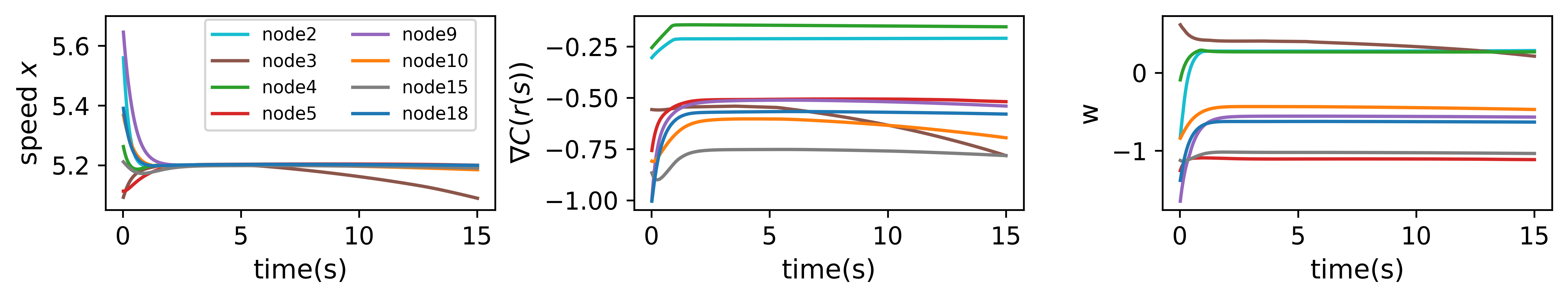}%
}
\caption{Dynamics of the system under four methods on 8 nodes with the required output agreement $\Bar{y}=5.2$. (a) NeuralPI-Comm achieves the output agreement level and identical marginal cost. (b) NeuralPI-WoComm achieves the output agreement but fails to converge to the identical-marginal-cost solution. 
(c) LinearPI-Comm is stable but has slower convergence compared with neural network-based approaches. (d) DenseNN-Comm leads to unstable behavior in node 3. }
\label{fig:Dynamic_Behavior_vehicle}
\end{figure}

Fig.~\ref{fig:Dynamic_Behavior_vehicle} shows the dynamics of velocity speed $\bm{x}$, marginal cost $\nabla\bm{C}(\bm{r}(\bm{s}))$ and external control action $\bm{w}$ on 8 nodes under the four methods. As guaranteed by Controller Design~\ref{design: optimal_flow}, NeuralPI-Comm in Fig.~\ref{fig:Dynamic_Behavior_vehicle}(a) reaches the same speed at 5.2 m/s and identical marginal cost, indicating that it achieves  the required output agreement level  with the lowest resource allocation cost. NeuralPI-WoComm in Fig.~\ref{fig:Dynamic_Behavior_vehicle}(b) also reaches the required output agreement level. However, the marginal cost converges at different levels for different nodes because of the lack of communication. LinearPI-Comm is stable and converges to the solution with identical marginal cost, but it has slower convergence compared with neural network-based approaches. 
DenseNN-Comm in Fig.~\ref{fig:Dynamic_Behavior_vehicle}(d) exhibits unstable behaviors in node 3.  
Therefore, it is necessary to enforce stability and steady-state optimality constraints on  controller design to provide performance guarantees.

\subsection{Power systems frequency control}
\label{subsec: exp_power}
Next we consider power systems where the frequencies of all generators need to be synchronized~\cite{sauer2017power, cui2022tps}.
Let $\mathcal{V}$ be the set of all the generators and $\mathcal{E}$ be the set of transmission lines. For each generator $i\in \mathcal{V}$, the rotating speed (i.e., frequency) $x_i$ changes with the mismatch between power generation and consumption. We aim to maintain the real-time balance of power generation and consumption by controlling the frequency of all generators at the same nominal value $\Bar{x}$ (e.g., 60Hz in the US). 

Let $P_i^m$ denote the fixed power generation  $d_i$ be the load at the node $i$.
The frequency dynamic of node $i\in\mathcal{V}$ is represented as~\cite{sauer2017power, cui2022tps} 
 \vspace{-0.2cm}
\begin{equation}\label{eq:power_node}
  \mathcal{V}_{i}: \quad\dot{x}_i=-\rho_i(x_i-\Bar{x})+P_i^m-d_i+u_i, \quad y_i=x_i.
 \end{equation}
where $\rho_i$ is the damping constant corresponding to the physical system. The input $ u_i=w_i-\sum_{l}E_{il}\mu_l$, where $w_i$  is the changes in the power generation  and $\mu_l$ is power flow along a  transmission line  $l$.
Note that  the system should withstand disturbances such as step load changes in  $d_i$. This is  quantified by limiting the maximum frequency deviation and quickly recovering back to the nominal frequency after disturbances.

By physical law
~\cite{sauer2017power, cui2022tps},  the active power flow in the line $l=(i,j)$  is $\mu_{l}=b_l\sin\left(\eta_{l}\right)$, where $\eta_l$ is the angle differences  that change with the difference of rotating speed  between node $i$ and $j$. Thus, the dynamics on an edge is
\vspace{-0.2cm}
\begin{equation}\label{eq:power_edge}
\mathcal{E}_{l}: \quad \dot{\eta}_{l}=\zeta_{l}, \quad \zeta_{l}=x_i-x_j,\quad \mu_{l}=b_l\sin\left(\eta_{l}\right),
\end{equation}
where  $b_l$ is the susceptance of the transmission line.
In vector form, we have $\bm{\zeta}=\bm{E}^{\top} \bm{y}$ and this recovers the input-output coupling of nodes and edges shown in Fig.~\ref{fig:network}.

Unlike the vehicle platooning problem, we cannot change the edge feedback function $\mu_{l}=b_l\sin\left(\eta_{l}\right)$ because it is determined by the physics. Here, we can only optimize $\bm{w}$ such that the frequency reaches the nominal value at the steady state and the cost of power generation is minimized.
Assumptions~\ref{ass: EIP_node}-\ref{ass:mapping} are verified in Appendix~\ref{app:assump_power}.

\subsubsection{Optimal resource allocation (economic dispatch)} Here we show an explicit derivation for the optimal resource allocation problem in~\eqref{eq: optimization_equivalent}. We aim to reach the output agreement such that $\Bar{y}=\Bar{x}$. Note that $\bm{h}^{-1}(\Bar{y}\mathbbold{1}_n)=\Bar{y}\mathbbold{1}_n$ and $k_{x,i}^{-1}(\Bar{y})=\rho_i(\Bar{x}-\Bar{x})+d_i-P_i^m=d_i-P_i^m$, we have $\bm{k_x}^{-1} (\bm{h}^{-1}(\Bar{y}\mathbbold{1}_n))=\bm{k_x}^{-1} (\Bar{y}\mathbbold{1}_n)=\bm{d}-\bm{P}^m$.

Then the constraint~\eqref{subeq:opt_equlity2} is written as $\bm{d} = \bm{w}^* -\bm{E}\bm{\mu}^*+\bm{P}^m$, which is the power balance equation.
The optimization problem~\eqref{eq: allocation_optimization}  is then written as
\vspace{-0.2cm}
\begin{subequations}\label{eq: optimization_equivalent_power}
\begin{align}
    \min_{\bm{w}^*, \bm{\mu}^*}& \sum_{i=1}^nC_i(w_i^*),\\
    \text{s.t. } & \bm{d} = \bm{w}^* -\bm{E}\bm{\mu}^*+\bm{P}^m\label{subeq:opt_equlity2_power}, 
\end{align}
\end{subequations}
which is the well-known economic dispatch problem in power systems that aims to serve demand with the lowest cost in power generation~\cite{DORFLER2017Auto, weitenberg2018robust}.

It is obvious that different $\bm{d}$ affects the constrains~\eqref{subeq:opt_equlity2_power} and thus changes the optimal solution to the resources allocation. Since the loads are time-varying, one important benefit of the proposed approach is that it distributedly attains the optimal solution  following the changes of the load levels without a centralized dispatch.

\subsubsection{Simulation setup}
We conduct experiments on the IEEE New England 10-machine 39-bus (NE39)
power network with parameters given in~\cite{athay1979practical, cui2022tps}.  We generate the training and test set of size 300 by randomly picking at most three generators to have a step load change uniformly distributed in $\texttt{uniform}[-1,1]\,\text{p.u.}$, where 1p.u.=100 MW is the base unit of power for the IEEE-NE39 test system.  Note that the load $d_i$ is a parameter in the node dynamics~\eqref{eq:power_node}. This experiment verifies the robustness of the controller under parameter changes. 
The state $\eta_l$ is initialized as 
the solution of power flow at the nominal frequency and $s_i$ is initialized as 0. The communication graph is randomly generated to be a regular graph with degree three.  
The episode number and batch size are 600 and 300,  respectively. The step-size in time is set as $\Delta t= 0.01s$ and  the number of time stages in a trajectory in the training set is $K=400$.

\subsubsection{Controller performances}
We implement external control law for  power output  $\bm{w}$ of generators to realize the agreement of frequency at 60Hz and reduce steady-state power generation cost. Apart from the accumulated frequency deviation, an important metric for the frequency control problem is the maximum frequency deviation (also known as the frequency nadir) after a disturbance~\cite{cui2022tps}. Hence, the transient cost is set to be $ J(\bm{y},\bm{w}) =\sum_{i=1}^n\big(\max_{k=1,\cdots,K}| y_i(k\Delta t)-\Bar{y}|+ 0.05\sum_{k=1}^{K}| y_i(k\Delta t)-\Bar{y}|+\sum_{k=1}^{K}c_i (w_i(k\Delta t))^4\big)$, where $c_i\sim \texttt{uniform}[0.25,0.75]$. The steady-state cost in resource allocation~\eqref{eq: allocation_optimization} is $C(\bm{w}) = \sum_{i=1}^n c_i (w_i^*)^4$, where the cost function is set as the power of four to demonstrate that the proposed approach is not restricted to quadratic cost functions. We use $w_i(30)$ to approximate $w_i^*$ since the dynamics approximately enter the steady state after $t=30s$ as we will show later in the simulation. 
The loss function in training is  $J(\bm{y},\bm{w})$, such that neural networks are optimized to reduce transient cost.

\begin{figure}[ht]
\centering
\subfloat[Training Loss]{\includegraphics[width=1.7
in]{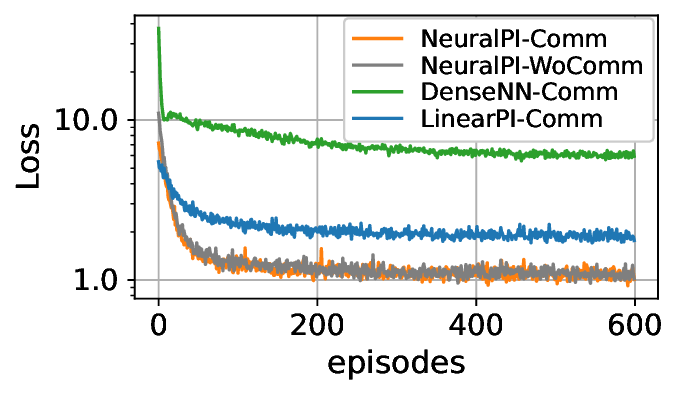}%
\label{fig_first_case_x}}
\hfil
\subfloat[Transient and steady cost ]{\includegraphics[width=1.7in]{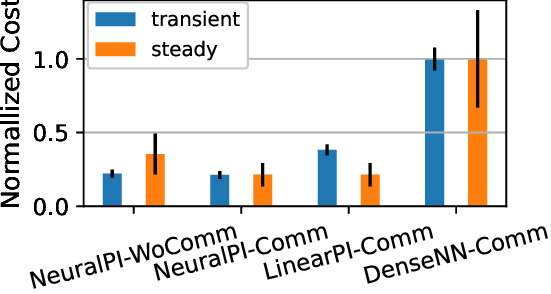}%
\label{fig_second_case_x}}
\caption{(a) Average batch loss along episodes. All converge, with the NeuralPI achieves the lowest cost. (b) The average transient cost and steady-state cost with error bar
on the randomly generated  test set with size 300. NeuralPI achieves  a transient cost that is much lower than others.
NeuralPI-Comm and LinearPI-Comm lead to the same lowest steady-state cost guaranteed by Controller Design~\ref{design: optimal_flow}.
 }
\label{fig:Loss_epi_power}
\end{figure}

Similar to the case study of the vehicle platoon, we compare the performance of four controllers. The average batch loss during episodes of training is shown in Fig.~\ref{fig:Loss_epi_power}(a). All of the four methods converge, with the NeuralPI achieving the lowest cost.
Fig.~\ref{fig:Loss_epi_power}(b) shows the transient and steady-state costs on the test set. 
NeuralPI achieves a transient cost that is much lower than the others. Note that the load changes lead to different solutions of optimal resource allocation problem~\eqref{eq: optimization_equivalent_power}, thus the steady-state cost also lies in a range. 
 Still, NeuralPI-Comm and LinearPI-Comm have the lowest possible steady-state cost, as guaranteed by Theorem~\ref{thm:equilibrium_cost}. 
 
\begin{figure}[t]
\centering
\subfloat[NeuralPI-Comm: dynamics of $\bm{x}$ , $\nabla\bm{C}(\bm{r}(\bm{s}))$  and $w$ \vspace{-0.2cm}
]{\includegraphics[width=3.4
in]{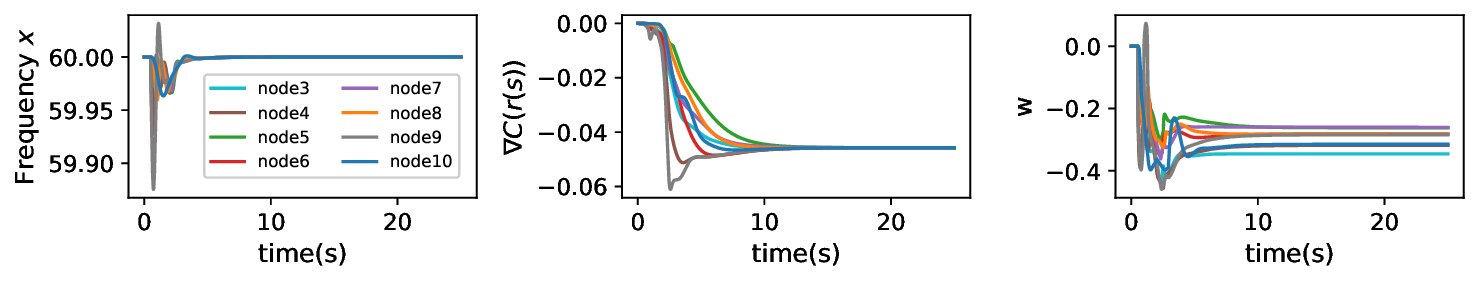}%
}
\hfil
\subfloat[NeuralPI-WoComm: dynamics of $\bm{x}$ , $\nabla\bm{C}(\bm{r}(\bm{s}))$ and $w$\vspace{-0.2cm}
]{\includegraphics[width=3.4
in]{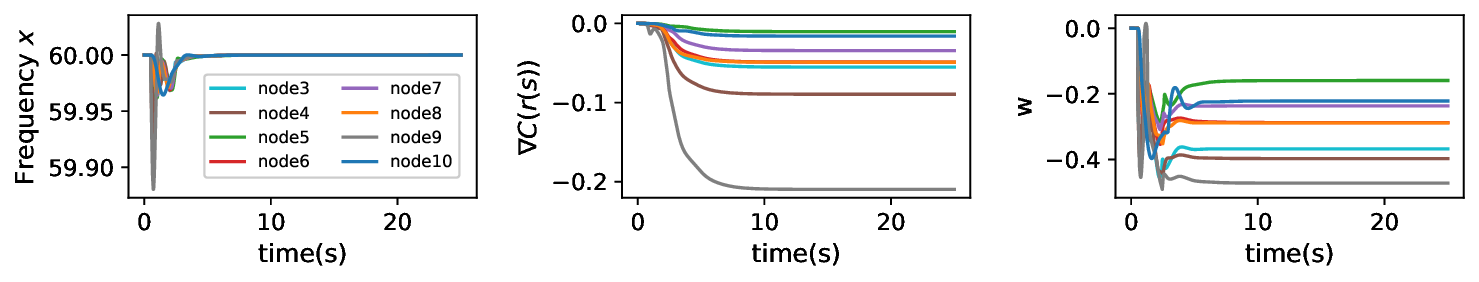}%
}
\hfil
\subfloat[LinearPI-Comm: dynamics of $\bm{x}$ , $\nabla\bm{C}(\bm{r}(\bm{s}))$and $w$\vspace{-0.2cm}
]{\includegraphics[width=3.4in]{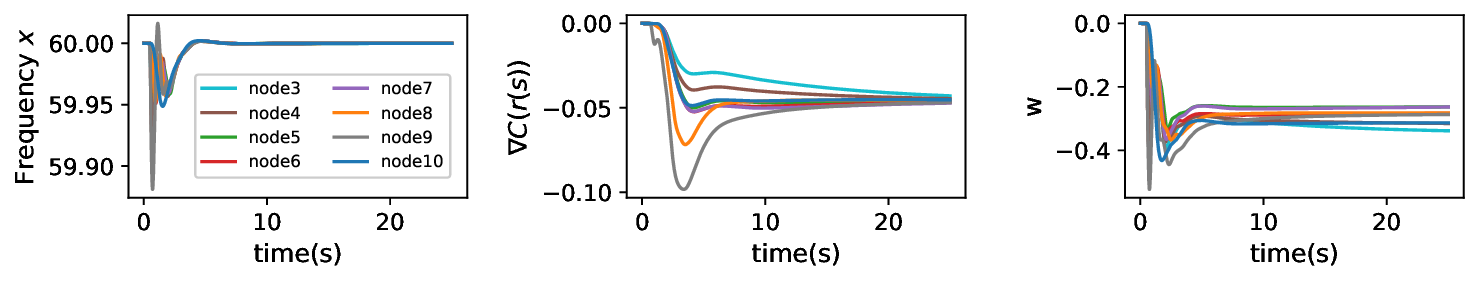}%
}
\hfil
\subfloat[DenseNN-Comm: dynamics of $\bm{x}$ , $\nabla\bm{C}(\bm{r}(\bm{s}))$ and $w$\vspace{-0.2cm}
]{\includegraphics[width=3.4in]{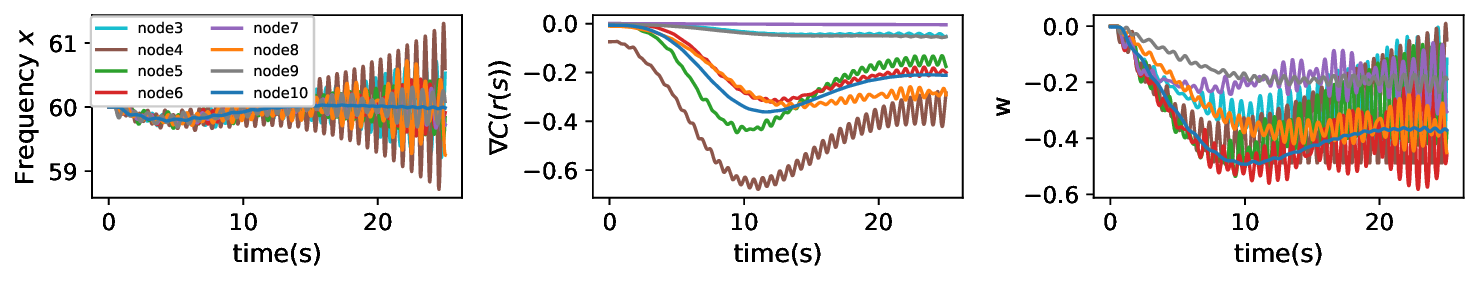}%
}
\caption{Dynamics of the system under four methods on 8 nodes with the required output agreement $\Bar{y}=60$ and  a step load change at 0.5s. (a) NeuralPI-Comm achieves the output agreement at 60Hz and identical marginal cost. (b) NeuralPI-WoComm achieves the output agreement but fails to converge to the identical-marginal-cost solution. (c) LinearPI-Comm is stable but has slower convergence compared with neural network-based approaches.
(d) DenseNN-Comm leads to large frequency deviations and oscillations. \vspace{-0.4cm}}
\label{fig:Dynamic_Behavior_power}
\end{figure}

With a step load change at 0.5s, Fig.~\ref{fig:Dynamic_Behavior_power} shows the dynamics of frequency $\bm{x}$, marginal cost $\nabla\bm{C}(\bm{r}(\bm{s}))$ and external control action $\bm{w}$ on 8 nodes under the four methods. As guaranteed by Controller Design~\ref{design: optimal_flow}, NeuralPI-Comm in Fig.~\ref{fig:Dynamic_Behavior_power}(a) achieves the output agreement at 60Hz and identical marginal cost, indicating that it achieves the lowest resource allocation cost. NeuralPI-WoComm in Fig.~\ref{fig:Dynamic_Behavior_power}(b) also reaches the output agreement at 60Hz guaranteed by Controller Design~\ref{design: agreement_set}. However, the marginal cost converges at different levels for different nodes because of the lack of communication. LinearPI-Comm in  Fig.~\ref{fig:Dynamic_Behavior_power}(c) converges to the solution with identical marginal cost, but the speed of convergence is slow. DenseNN-Comm in   Fig.~\ref{fig:Dynamic_Behavior_power}(d)  exhibits unstable behavior with large oscillations.   Hence, 
the guarantees provided in Controller Design~\ref{design: agreement_set} and Controller Design~\ref{design: optimal_flow} are robust to parameter changes, which have significant practical importance. Controller Design~\ref{design: optimal_flow} further realizes the economic dispatch of generators under different load levels distributedly.



\section{Conclusion} \label{sec:conclusion}
This paper proposes structured Neural-PI controllers for networked systems where node dynamics are equilibrium independent passive. The proposed controllers have provable guarantees on stability and can distributedly achieve optimal resource allocation at the steady state. Experiments demonstrate that the proposed approach can improve both transient and steady-state performances and is also robust to parameter changes, while unstructured neural networks lead to unstable behaviors. Important future directions include relaxing the limitations on the one-dimensional SISO nodal system and incorporating other safety constraints. 
\bibliography{ref.bib}

\appendices

\vspace{-0.4cm}

\subsection{Proof of Lemma~\ref{lem:sign-cross}}
\label{app:lem_sign-cross}
\begin{proof}
We start by showing that
$\bm{r}(\bm{s})^T \hat{\bm{c}}\tilde{\bm{E}}\bm{\phi}\left(\tilde{\bm{E}}^\top\nabla \bm{C}\left(\bm{r}\left(\bm{s}\right)\right)\right)\geq0$
with equality holds if and only if $\nabla \bm{C}(\bm{r}(\bm{s}))\in\range{\mathbbold{1}_n}$.
Expanding the left side of~\eqref{eq:cross_communication} and pre-multiplying $\bm{r}\left(\bm{s}\right)$  gives
\begin{equation*}\label{eq: pcEPED}
\begin{split}
     &  (\bm{r}\left(\bm{s}\right))^\top \left(\hat{\bm{c}} \tilde{\bm{E}} \bm{\phi}\left(\tilde{\bm{E}}^\top\nabla \bm{C}(\bm{r}\left(\bm{s}\right))\right)\right)
     \\
    &= \!\sum_{ l=(\!i,j\!)\in  \tilde{\mathcal{E}} }\!\phi_{l}\!\left(\nabla C_i\!\left(r_i(s_i\!)\right)\!-\!\nabla C_j\!\left(r_j(s_j\!)\right)\right)\!\cdot\!\left(c_ir_i(s_i\!)\!-\!c_jr_j(s_j\!)\right) 
    \\
    &=\sum_{ l=(i,j)\in  \tilde{\mathcal{E}} }\phi_{l}\left(\nabla  C_\mathrm{o}\left(c_ir_i(s_i)\right)-\nabla  C_\mathrm{o}\left(c_jr_j(s_j)\right)\right)\\
    &\hspace{1.5cm}\cdot\left(c_ir_i(s_i)-c_jr_j(s_j)\right) 
\end{split}
\end{equation*}
where the last step follows from the cost function in Assumption~\ref{ass: cost-grad-scale} that $\nabla C_i(r_i(s_i))=\nabla  C_\mathrm{o}(c_ir_i(s_i))$ for all $i\in\mathcal{V}$.

Since $ C_\mathrm{o}(\cdot)$ is strictly convex, its gradient $\nabla  C_\mathrm{o}(\cdot)$ is strictly increasing~\cite{Boyd2004convex}. Thus, 
\begin{equation}\label{eq: app_cross_positive}
\!\left(\nabla  \!C_\mathrm{o}\!\left(c_ir_i(s_i\!)\!\right)\!-\!\nabla  C_\mathrm{o}\!\left(c_jr_j(s_j\!)\!\right)\right)\!\left(c_ir_i(s_i\!)\!-\!c_jr_j(s_j\!)\right)\! \geq \!0    
\end{equation}
with equality holds if and only if $\nabla  C_\mathrm{o}\left(c_ir_i(s_i)\right)=\nabla  C_\mathrm{o}\left(c_jr_j(s_j)\right)$.

 By Controller Design~\ref{design: optimal_flow}, $\phi_{l}\left(\nabla  C_\mathrm{o}(c_ir_i(s_i))\!-\!\nabla  C_\mathrm{o}(c_jr_j(s_j))\right)$ is the same sign with $\nabla  C_\mathrm{o}(c_ir_i(s_i))\!-\!\nabla  C_\mathrm{o}(c_jr_j(s_j))$. Hence,~\eqref{eq: app_cross_positive} implies
$$ 
\phi_{l}\left(\nabla  C_\mathrm{o}\!\left(c_ir_i(s_i)\right)\!-\!\nabla  C_\mathrm{o}\left(c_jr_j(s_j)\right)\right)\left(c_ir_i(s_i)\!-\!c_jr_j(s_j)\right)\! \geq\! 0
$$ 
with equality holds if and only if $\nabla  C_\mathrm{o}\!\left(c_ir_i(s_i)\right)=\nabla  C_\mathrm{o}\!\left(c_jr_j(s_j)\right)$. This implies $\nabla C_i(r_i(s_i))=\nabla C_j(r_j(s_j))$ $ \forall l=(i,j)\in  \tilde{\mathcal{E}}$ for cost functions satisfying Assumption~\ref{ass: cost-grad-scale}.
Since the graph is connected, we further have $\nabla C_1(r_1(s_1))=\!\ldots\!=\!\nabla C_n(r_n(s_n))$, i.e., $\nabla \bm{C}(\bm{r}(\boldsymbol{s}))\in\range{\mathbbold{1}_n}$.

Then we prove that $\hat{\bm{c}}\tilde{\bm{E}}\bm{\phi}\left(\tilde{\bm{E}}^\top\nabla \bm{C}\left(\bm{r}\left(\bm{s}\right)\right)\right)=\mathbbold{0}_n$ if and only if $\nabla \bm{C}(\bm{r}(\bm{s}))\in\range{\mathbbold{1}_n}$ by showing sufficiency and necessity. 
 If $\nabla C(\bm{r}(\bm{s}))\in\range{\mathbbold{1}_n}$, we have $\tilde{\bm{E}}^\top\nabla \bm{C}\left(\bm{r}\left(\bm{s}\right)\right)=\mathbbold{0}_n$ and thus $\hat{\bm{c}}\tilde{\bm{E}}\bm{\phi}\left(\tilde{\bm{E}}^\top\nabla \bm{C}\left(\bm{r}\left(\bm{s}\right)\right)\right)=\mathbbold{0}_n$. On the other hand, if $\hat{\bm{c}}\tilde{\bm{E}}\bm{\phi}\left(\tilde{\bm{E}}^\top\nabla \bm{C}\left(\bm{r}\left(\bm{s}\right)\right)\right)=\mathbbold{0}_n$, Multiplying both sides by $(\bm{r}\left(\bm{s}\right))^\top$ gives  $\bm{r}(\bm{s})^\top \hat{\bm{c}}\tilde{\bm{E}}\bm{\phi}\left(\tilde{\bm{E}}^\top\nabla \bm{C}\left(\bm{r}\left(\bm{s}\right)\right)\right)=0$ and therefore  $\nabla C(\bm{r}(\bm{s}))\in\range{\mathbbold{1}_n}$. Hence, $\hat{\bm{c}}\tilde{\bm{E}}\bm{\phi}\left(\tilde{\bm{E}}^\top\nabla \bm{C}\left(\bm{r}\left(\bm{s}\right)\right)\right)=\mathbbold{0}_n$ if and only if $\nabla C(\bm{r}(\bm{s}))\in\range{\mathbbold{1}_n}$.
\end{proof}

\subsection{Proof of Theorem~\ref{thm:equilibrium_cost}}\label{app:thm_equilibrium_cost}
\begin{proof}
At the equilibrium, we have $\bm{f}(\bm{x}^*,\bm{u}^*)=\bm{0}$ and  $\bm{\zeta}^*=\bm{0}$. For a connected graph, the null space of $\bm{E}^\top$ is $\mathcal{N}(\bm{E}^\top)=\text{range}\left\{\mathbbold{1}_n\right\}$~\cite{biggs1993algebraic}.
Using $\bm{\zeta}^*=\bm{E}^\top \bm{y}^*=\bm{0}$, we have $\bm{y}^*=\hat{y}\mathbbold{1}_n$. The right side of~\eqref{subeq:s_opf} equals to zero at the equilibrium gives $(\hat{y}\mathbbold{1}_n-\Bar{y}\mathbbold{1}_n)=-\hat{\bm{c}} \bm{E} \bm{\phi}\left(\bm{E}^\top\nabla \bm{C}(\bm{r}(\bm{s}))\right)$. Multiplying both sides by $\mathbbold{1}_n^\top\hat{\bm{c}}^{-1}$ yields $(\hat{y}-\Bar{y})\mathbbold{1}_n^\top\hat{\bm{c}}^{-1}\mathbbold{1}_n =-\mathbbold{1}_n^\top \bm{E} \bm{\phi}\left(\bm{E}^\top\nabla \bm{C}(\bm{r}(\bm{s}))\right)$, which equals to zero since $\mathbbold{1}_n^\top \bm{E}= \mathbbold{0}_n$ for a connected graph.
This implies $\hat{y}=\Bar{y}$ since $\mathbbold{1}_n^T\hat{\bm{c}}^{-1}\mathbbold{1}_n>0$ for $\hat{\bm{c}}\succ 0$. Therefore, $\bm{y}^*=\ \Bar{y}\mathbbold{1}_n$ and thus $\bm{x}^*=\ \bm{h}^{-1}\left(\Bar{y}\mathbbold{1}_n\right)$ by bijective mapping of $\bm{h}(\cdot)$.

Moreover, $\bm{y}^*=\ \Bar{y}\mathbbold{1}_n$ implies $\hat{\bm{c}} \tilde{\bm{E}} \bm{\phi}\left(\tilde{\bm{E}}^\top\nabla \bm{C}(\bm{r}\left(\bm{s}^*\right))\right)=\mathbbold{0}_n$. By Lemma~\ref{lem:sign-cross}, $\nabla \bm{C}(\bm{r}(\boldsymbol{s}))\in\range{\mathbbold{1}_n}$ and thus there exists a scalar $\gamma$ such that  $\nabla C_i(r_i\left(s_i^*\right))= \gamma$ for all $i\in\mathcal{V}$. This implies  $\nabla  C_\mathrm{o}(c_ir_i\left(s_i^*\right))= \gamma$ by Assumption~\ref{ass: cost-grad-scale}. The strict convexity of $C_\mathrm{o}(\cdot)$ implies that $\nabla C_\mathrm{o}(\cdot)$ is a strictly increasing function, which guarantees the existence of $\nabla C_\mathrm{o}^{-1}(\cdot)$ that is also a strictly increasing function.
Hence, $r_i(s_i^*)=\nabla C_\mathrm{o}^{-1}(\gamma)c_i^{-1}$ and compactly we have $\bm{r}\left(\bm{s}^*\right)=\nabla C_\mathrm{o}^{-1}(\gamma)\hat{\bm{c}}^{-1}\mathbbold{1}_n$.


From the bijective mapping of $\bm{k_x}(\cdot)$, $\bm{u}^* =\bm{k_x}^{-1}(\bm{x}^*)= \bm{k_x}^{-1} (\bm{h}^{-1}(\Bar{y}\mathbbold{1}_n)).$ 
From $\bm{p}(-\bm{y}^*+\Bar{y}\mathbbold{1}_n) = \mathbbold{0}_n$, we have
$\bm{u}^* =-\bm{E}\bm{\mu}^*+ \bm{r}\left(\bm{s}^*\right) $ and therefore $-\bm{E}\bm{\mu}^* = \bm{k_x}^{-1} (\bm{h}^{-1}(\Bar{y}\mathbbold{1}_n)) - \nabla C_\mathrm{o}^{-1}(\gamma)\hat{\bm{c}}^{-1}\mathbbold{1}_n$. 
Since  $\mathbbold{1}_n^\top \bm{E} \bm{\mu}^*=0$, we have  $\nabla C_\mathrm{o}^{-1}(\gamma)=-\left(\mathbbold{1}_n^\top\bm{k_x}^{-1} (\bm{h}^{-1}(\Bar{y}\mathbbold{1}_n)) \right)/\left(\sum_{i=1}^n c_i^{-1}\right)$.  The uniqueness of $\gamma$ is guaranteed by the strict increasing property of function $\nabla C_\mathrm{o}^{-1}(\gamma)$. Similarly, the uniqueness of $\bm{s}^*$ satisfying $\bm{r}\left(\bm{s}^*\right)=\nabla C_\mathrm{o}^{-1}(\gamma)\hat{\bm{c}}^{-1}\mathbbold{1}_n$ is guaranteed by the strictly increasing property of function $r_i(\cdot)$ for $i\in\mathcal{V}$.

Then, we prove the uniqueness of $\bm{\eta}^*$ by contradiction. By $\bm{\mu}^*=\bm{\psi}(\bm{\eta}^*)$, we have $-\bm{E}\bm{\psi}(\bm{\eta}^*)=\bm{k_x}^{-1} (\bm{h}^{-1}(\Bar{y}\mathbbold{1}_n))-\nabla C_\mathrm{o}^{-1}(\gamma)\hat{\bm{c}}^{-1}\mathbbold{1}_n$. 
Suppose there is $\hat{\bm{\eta}}\in\real^m$ and $\hat{\bm{\eta}} \neq \bm{\eta}^*$ such that $\bm{E}\bm{\psi}(\bm{\eta}^*)=\bm{E}\bm{\psi}(\hat{\bm{\eta}})$. 
Since $\bm{\eta}\in\mathcal{R}(\bm{E}^\top)$, there exist $\bm{z}^*\in\real^n$ and $\hat{\bm{z}}\in\real^n$ such that $\bm{\eta}^*=\bm{E}^\top\bm{z}^*$ and $\hat{\bm{\eta}}=\bm{E}^\top\hat{\bm{z}}$. 
Then $(\bm{z}^*-\hat{\bm{z}})^\top \left(\bm{E}\bm{\psi}(\bm{\eta}^*)-\bm{E}\bm{\psi}(\hat{\bm{\eta}}) \right)
=
(\bm{z}^*-\hat{\bm{z}})^\top \left(\bm{E}\bm{\psi}(\bm{E}^\top\bm{z}^*)-\bm{E}\bm{\psi}(\bm{E}^\top\hat{\bm{z}}) \right)
=
\sum_{l=(i,j)\in\mathcal{E}}(z^*_{ij}-\hat{z}_{ij})\left(\psi_l(z^*_{ij})-\psi_l(\hat{z}_{ij})\right)$. Since $\psi_l(\cdot)$ is monotonically increasing, we have $(z^*_{ij}-\hat{z}_{ij})\left(\psi_l(z^*_{ij})-\psi_l(\hat{z}_{ij})\right)\geq 0$ for all $l$ with equality only holds when $z^*_{ij}=\hat{z}_{ij}$ (which is equivalent to $\bm{E}^\top\bm{z}^*=\bm{E}^\top\hat{\bm{z}}$ and thus $\bm{\eta}^*=\hat{\bm{\eta}}$). Hence, $\bm{E}\bm{\psi}(\bm{\eta}^*)=\bm{E}\bm{\psi}(\hat{\bm{\eta}})$ in and only if $\bm{\eta}^*=\hat{\bm{\eta}}$.
\end{proof}

\subsection{Proof of Theorem~\ref{thm:stack_relu}}\label{app:lemma_stack_relu }
We start by showing that the design of the stacked-ReLU neural network forms a piece-wise linear function that is strictly increasing and across the origin. 
Expanding the terms in  the stacked-ReLU neural network gives
\vspace{-0.2cm}
\begin{equation*}\label{eq:stacked-relu_expand}
\begin{split}
     g(z)=& \sum_{j=1}^d\alpha_j^+\sigma(z-\beta_j^+)
     +\sum_{j=1}^d\alpha_j^-\sigma(-z+\beta_j^-)
     \\
     \mbox{where }
     &-\infty<\sum_{j=1}^{d'} \alpha_j^-<0<\sum_{j=1}^{d'} \alpha_j^+<\infty \,, \forall d' = 1, ..., d\\
     &\beta_d^-\leq\cdots\leq \beta_1^-=0=\beta_1^+\leq\cdots\leq \beta_d^+.
\end{split}
\end{equation*}

Note that the neuron $\alpha_j^+\sigma(z-\beta_j^+)=\alpha_j^+(z-\beta_j^+)$ if $z\geq\beta_j^+$ (sometimes called activated) and equals to zero otherwise. Similarly, the  neuron $\alpha_j^-\sigma(-z+\beta_j^-)=\alpha_j^-(-z+\beta_j^-)$ if $z\leq\beta_j^-$ and equals to zero otherwise. Hence, the constraint $\beta_d^-\leq\cdots\leq \beta_1^-=0=\beta_1^+\leq\cdots\leq \beta_d^+$ guarantees that $g(0)=0$ and the neurons activate in sequence such that
\begin{equation*}
\begin{split}
 &g(z)\\
&= \left\{\begin{matrix*}[l]
\sum_{j=1}^d\alpha_j^+\left(z-\beta_j^+\right), \, z> \beta_{d}^+
\\[0.8em]
\sum_{j=1}^k\alpha_j^+\left(z-\beta_j^+\right), \, z\in (\beta_{k}^+, \beta_{k+1}^+], k=1,\cdots,d-1 \\[0.8em]
\sum_{j=1}^k\alpha_j^-\left(-z\!+\!\beta_j^-\right), \, z\in [\beta_{k+1}^-, \beta_k^-), k=1,\cdots\!,d\!-\!1 \\[0.8em]
\sum_{j=1}^d\alpha_j^-\left(-z+\beta_j^-\right), \, z<\beta_d^-.
\end{matrix*}\right.   
\end{split}
\end{equation*}
Hence, $g(z)$ forms a piece-wise linear function across the origin, and constraints $-\infty<\sum_{j=1}^{d'} \alpha_j^-<0<\sum_{j=1}^{d'} \alpha_j^+<\infty, \forall d' = 1, ..., d$ further guarantee that the slope is positive, i.e., the function is strictly increasing.

The proof of the universal approximation of monotonic functions follows in two step. First, we show that these functions can be approximated with arbitrary small error by  piece-wise linear functions. Next, we show that the piece-wise linear function can be constructed exactly using the stacked ReLU structure.   
Let $\mathcal{Z}$ be a closed interval in $\real$ and $r(z):\mathcal{Z}\mapsto\real$ be a bounded, $L$-Lipschitz continuous and monotonically increasing function through the origin. Define an equally spaced grid of points on $\mathcal{Z}$, where $ \tau  =\frac{1}{d}$ is the spacing between grid points along each dimension. Corresponding to each grid interval $[(k-1) \tau  ,k \tau  ]$ with $k=1,\cdots,d$, assign a linear function
\begin{equation}\label{eq:piece_wise_linear}
\tilde{g}(z)=r((k-1)\tau  )+\frac{r(k\tau  )-r((k-1)\tau  )}{ \tau  }(z-(k-1)\tau  ),    
\end{equation}
where $\tilde{g}((k-1)\tau  )=r((k-1)\tau  )$ and $\tilde{g}(k\tau  )=r(k\tau  )$.

Since $r(\cdot)$ is monotonically increasing, we have $r((k-1)\tau  )\leq r(z) \leq r(k\tau  )$ and $r((k-1)\tau  )\leq \tilde{g}(z) \leq r(k\tau  )$ for all $z\in [(k-1)\tau  ,k\tau  ]$. The approximation error is bounded by
\begin{equation}\label{eq:Bound_relu}
    |\tilde{g}(z)-r(z)|\leq |r(k\tau  )-r((k-1)\tau  )|\leq L \tau,
\end{equation}
where the last inequality follows $r(\cdot)$ is $L$-Lipschitz.

Without loss of generosity, assume that $z\geq0$ and thus the function~\eqref{eq:stacked-relu} is reduced to $g(z)=\sum_{j=1}^d\alpha_j^+\sigma(z-\beta_j^+)$. Let 
$\beta_k^{+}=(k-1) \tau  $, $\sum_{j=1}^{k} \alpha_j^{+}=\frac{r(k \tau  )-r((k-1) \tau  )}{ \tau  }$ for $k=1,2,\cdots,d.$ Then the construction of $g(z)$ is exactly the same as $\tilde{g}(z)$. Therefore,  $|g(z)-r(z)|$ can also be bounded
by $L\tau$ using~\eqref{eq:Bound_relu}. We take $\tau <\frac{\epsilon}{L}$ to complete the proof.

\subsection{Verification of assumptions for vehicle platooning}\label{app:assump_vehicle} 

We check assumptions~\ref{ass: EIP_node}-\ref{ass:mapping} for this networked system. 

\textit{Well defined bijective mapping.} For the node dynamics~\eqref{eq:traffic_dyn_node}, $f_i(x_i,u_i)=\kappa_{i}\left(-(x_{i}-\lambda_{i}^{0})+\frac{1}{\rho_i}u_{i}\right)$ and $h_i(x_i)=x_i$, where $h_i(\cdot)$ is bijective.
    At the equilibrium, $f_i(x_i^*,u_i^*)=0$ gives  $-(x_{i}^*-\lambda_{i}^{0})+\frac{1}{\rho_i}u_{i}^*=0$.
    This yields a well-defined bijective mapping  $k_{x,i}(u_i^*)=\frac{1}{\rho_i}u_i^*+\lambda_i^0$, $k_{y,i}(u_i^*)=\frac{1}{\rho_i}u_i^*+\lambda_i^0$, with inverses $h_i(\cdot)$ and $k_{x,i}(\cdot)$ are $h_i^{-1}(y_i^*) = y_i^*$ and $k_{x,i}^{-1}(x_i^*)=\rho_i(x_i^*-\lambda_i^0)$, respectively.

    \textit{Strict EIP of node dynamics. }
    The well defined bijective mapping $k_{x,i}(\cdot)$ guarantees that for every $u_i^*\in \mathcal{U}_i^*$, there exists a unique $x_i^*\in\mathcal{X}$ such that $f_i(x_i^*, u_i^*)=0$. Let the storage function be $W_{i}^\mathcal{V}\left(x_{i}, x_{i}^*\right) = \frac{\rho_i}{2\kappa_{i}}(x_i-x_i^*)^2$. Then 
    \begin{equation*}
        \begin{split}
        W_{i}^\mathcal{V}\left(x_{i}, x_{i}^*\right)
        &=\rho_i(x_i-x_i^*)\dot{x_i}\\
        &=\rho_i(x_i-x_i^*)\left(-(x_{i}-\lambda_{i}^{0})+\frac{1}{\rho_i} u_{i}\right)\\
        &\stackrel{\circled{1}}{=}-\rho_i(x_i-x_i^*)^2+(x_i-x_i^*)(u_i-u_i^*)\\
        &\stackrel{\circled{2}}{=}-\rho_i(y_i-y_i^*)^2+(y_i-y_i^*)(u_i-u_i^*) 
        \end{split}
    \end{equation*}
    where $\circled{1}$ follows from $-(x_{i}^*-\lambda_{i}^{0})+\frac{1}{\rho_i}u_{i}^*=0$ and  $\circled{2}$ follows from $y_i=x_i$ by definition. Hence, each node dynamics~\eqref{eq:traffic_dyn_node} is strictly EIP with the storage function $W_{i}^\mathcal{V}\left(x_{i}, x_{i}^*\right)$.

\subsection{Verification of assumptions for frequency control}\label{app:assump_power} 

We check assumptions~\ref{ass: EIP_node}-\ref{ass:mapping} for this networked system. 

\textit{Well defined bijective mapping.} 
For the node dynamics~\eqref{eq:traffic_dyn_node}, $f_i(x_i,u_i)=-\rho_i(x_i-\Bar{x})+P_i^m-d_i+u_i$ and $h_i(x_i)=x_i$, where $h_i(\cdot)$ is obviously bijective. At the equilibrium, $f_i(x_i^*,u_i^*)=0$ gives  $-\rho_i(x_i^*-\Bar{x})+P_i^m-d_i+u_i^*=0$.
    This yields a well-defined bijective mapping  $k_{x,i}(u_i^*)=\Bar{x}+(P_i^m-d_i+u_i^*)/\rho_i$, $k_{y,i}(u_i^*)=\Bar{x}+(P_i^m-d_i+u_i^*)/\rho_i$. The corresponding inverse function of $h_i(\cdot)$ and $k_{x,i}(\cdot)$ are $h_i^{-1}(y_i^*) = y_i^*$ and $k_{x,i}^{-1}(x_i^*)=\rho_i(x_i^*-\Bar{x})+d_i-P_i^m$, respectively.

    \textit{Strict EIP of node dynamics. }
    The well defined bijective mapping $k_{x,i}(\cdot)$ guarantees that for every equilibrium $u_i^*\in \mathcal{U}_i^*$, there exists a unique $x_i^*\in\mathcal{X}$ such that $f_i(x_i^*, u_i^*)=0$. Let the storage function be $W_{i}^\mathcal{V}\left(x_{i}, x_{i}^*\right) = \frac{1}{2}(x_i-x_i^*)^2$. Then 
    \begin{equation*}
        \begin{split}
        \dot{W}_{i}^\mathcal{V}\left(x_{i}, x_{i}^*\right)
        &=(x_i-x_i^*)\dot{x_i}\\
        &=(x_i-x_i^*)\left(-\rho_i(x_i-\Bar{x})+P_i^m-d_i+u_i\right)\\
        &\stackrel{\circled{1}}{=}-\rho_i(x_i-x_i^*)^2+(x_i-x_i^*)(u_i-u_i^*)\\
        &\stackrel{\circled{2}}{=}-\rho_i(y_i-y_i^*)^2+(y_i-y_i^*)(u_i-u_i^*) 
        \end{split}
    \end{equation*}
    where $\circled{1}$ follows from the equilibrium $\left(-\rho_i(x_i^*-\Bar{x})+P_i^m-d_i+u_i^*\right)=0$ and  $\circled{2}$ follows from $y_i=x_i$ by definition. Since $\rho_i>0$, each node dynamics~\eqref{eq:traffic_dyn_node} is strictly EIP with $\rho_i$ and the storage function $W_{i}^\mathcal{V}\left(x_{i}, x_{i}^*\right)$

\textit{Strictly increasing of edge feedback functions.} 
We adopt a common assumption in literature that the power system operates with angle differences $\eta_l$ in the range $\mathcal{H}=\left\{\bm{\eta}|\eta_l\in (-\pi/2,\pi/2) \, \forall l \in \mathcal{E}\right\}$, which is sufficiently large to include almost all practical scenarios~\cite{weitenberg2018robust, DORFLER2017Auto, sauer2017power}. Since $\sin(\eta_l)$ is strictly monotonically increasing in  $(-\pi/2,\pi/2)$, the conditions on strictly increasing of edge feedback functions are satisfied.

\begin{IEEEbiography}[{\includegraphics[width=1in,height=1.25in,clip,keepaspectratio]{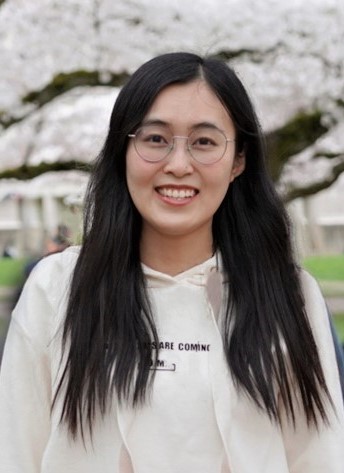}}]{Wenqi Cui}
received the B.Eng. degree and M.S. degree in electrical engineering from Southeast University, Nanjing, China, and Zhejiang University, Hangzhou, China, in 2016 and 2019, respectively. She is currently working toward the Ph.D. degree in Electrical Engineering at the University of Washington, Seattle, WA, USA. She works on control, optimization, and machine learning, with applications in power systems. \vspace{-0.5cm}
\end{IEEEbiography}

\begin{IEEEbiography}
[{\includegraphics[width=1in,height=1.25in,clip,keepaspectratio]{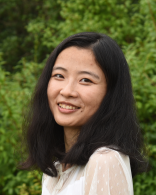}}]%
{Yan Jiang} received the B.Eng. degree in electrical engineering and automation from Harbin Institute of Technology, Harbin, CHN, in 2013, the M.S. degree in electrical engineering from Huazhong University of Science and Technology, Wuhan, CHN, in 2016, and the Ph.D. degree in electrical engineering with the M.S.E. degree in Applied Mathematics and Statistics from Johns Hopkins University, Baltimore, USA, in 2021. She is currently a Postdoctoral Scholar with the Department of Electrical and Computer Engineering at University of Washington, Seattle, USA. Her research interests lie in the area of control of power systems.\vspace{-0.5cm}
\end{IEEEbiography}

\begin{IEEEbiography}
	[	{\includegraphics[width=1in,height=1.25in,clip]{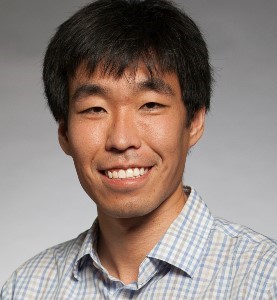}}
	] {Baosen Zhang} received his Bachelor of Applied Science in Engineering Science degree from the University of Toronto in 2008; and his PhD degree in Electrical Engineering and Computer Sciences from University of California, Berkeley in 2013. He was a Postdoctoral Scholar at Stanford University. He is currently an Associated Professor in Electrical and Computer Engineering at the University of Washington, Seattle, WA. His research interests are in control, optimization and learning applied to power systems and other cyberphysical systems. He received the NSF CAREER award as well as several best paper awards. \vspace{-0.5cm} \end{IEEEbiography}

 \begin{IEEEbiography}
	[	{\includegraphics[width=1in,height=1.25in,clip]{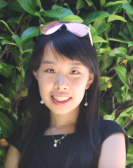}}
	] {Yuanyuan Shi} is an Assistant Professor of Electrical and Computer Engineering at the University of California, San Diego. She received her Ph.D. in Electrical Engineering, masters in Electrical Engineering and Statistics, all from the University of Washington, in 2020. From 2020 to 2021, she was a postdoctoral scholar at the California Institute of Technology. Her research interests include machine learning, dynamical systems, and control, with applications to sustainable power and energy systems. \end{IEEEbiography}
\end{document}